\documentclass[journal]{IEEEtran}
\usepackage{cite}
\usepackage{graphicx}
\usepackage{amsfonts}
\usepackage{amssymb}
\usepackage{amsmath}
\usepackage{mathrsfs}
\usepackage{bm}
\usepackage{amsmath}
\usepackage{mdwmath}
\usepackage{mdwtab}
\newtheorem{remark}{Remark}
\newtheorem{corollary}{Corollary}
\newcommand\smallO[1]{
  \mathchoice
  {
    \mathop{}\mathopen{}{\scriptstyle\mathcal{O}}\mathopen{}\left(#1\right)
  }
  {
    \mathop{}\mathopen{}{\scriptstyle\mathcal{O}}\mathopen{}\left(#1\right)
  }
  {
    \mathop{}\mathopen{}{\scriptscriptstyle\mathcal{O}}\mathopen{}\left(#1\right)
  }
  {
    \mathop{}\mathopen{}{o}\mathopen{}\left(#1\right)
  }
}

\begin{document}
\title{Dynamic Channel Acquisition in MU-MIMO}
\author{Zhiyuan Jiang,~\IEEEmembership{Student Member,~IEEE}, Sheng Zhou,~\IEEEmembership{Member,~IEEE}, and Zhisheng Niu,~\IEEEmembership{Fellow,~IEEE}
\thanks{The authors are with Tsinghua National Laboratory for Information Science and Technology, Tsinghua University, Beijing 100084, China. Email:
jiang-zy10@mails.tsinghua.edu.cn, \{sheng.zhou,niuzhs\}@tsinghua.edu.cn.

This work is sponsored in part by the National Basic Research Program of China (973 Program: 2012CB316001), the National Science Foundation of China (NSFC) under grant No. 61201191 and No. 61321061, and Hitachi R\&D Headquarter.}}
\maketitle
\vspace{-1cm}
\begin{abstract}
Multiuser multiple-input-multiple-output (MU-MIMO) systems are known to be hindered by \emph{dimensionality loss} due to channel state information (CSI) acquisition overhead. In this paper, we investigate user-scheduling in MU-MIMO systems on account of CSI acquisition overhead, where a base station dynamically acquires user channels to avoid choking the system with CSI overhead. The \emph{genie-aided optimization problem} (GAP) is first formulated to maximize the Lyapunov-drift every scheduling step, incorporating user queue information and taking channel fluctuations into consideration. The scheduling scheme based on GAP, namely the GAP-rule, is proved to be throughput-optimal but practically infeasible, and thus serves as a performance bound. In view of the implementation overhead and delay unfairness of the GAP-rule, the $T$-frame dynamic channel acquisition scheme and the power-law DCA scheme are further proposed to mitigate the implementation overhead and delay unfairness, respectively. Both schemes are based on the GAP-rule and proved throughput-optimal. To make the schemes practically feasible, we then propose the heuristic schemes, \emph{queue-based quantized-block-length user scheduling scheme} (QQS), $T$-frame QQS, and power-law QQS, which are the practical versions of the aforementioned GAP-based schemes, respectively. The QQS-based schemes substantially decrease the complexity, and also perform fairly close to the optimum. Numerical results evaluate the proposed schemes under various system parameters.
\end{abstract}
\section{Introduction}
Multiuser multiple-input-multiple-output (MU-MIMO) technology enables simultaneous (on the same time-frequency resource) data transmissions to a multiplicity of autonomous terminals via distinguishable spatial modes. With perfect channel state information at transmitter (CSIT) and at receiver (CSIR), the capacity of the MU-MIMO system is significantly larger than that of the system without spatial multiplexing transmissions \cite{Caire10}.

\begin{figure}[!t]
\centering
\includegraphics[width=0.4\textwidth]{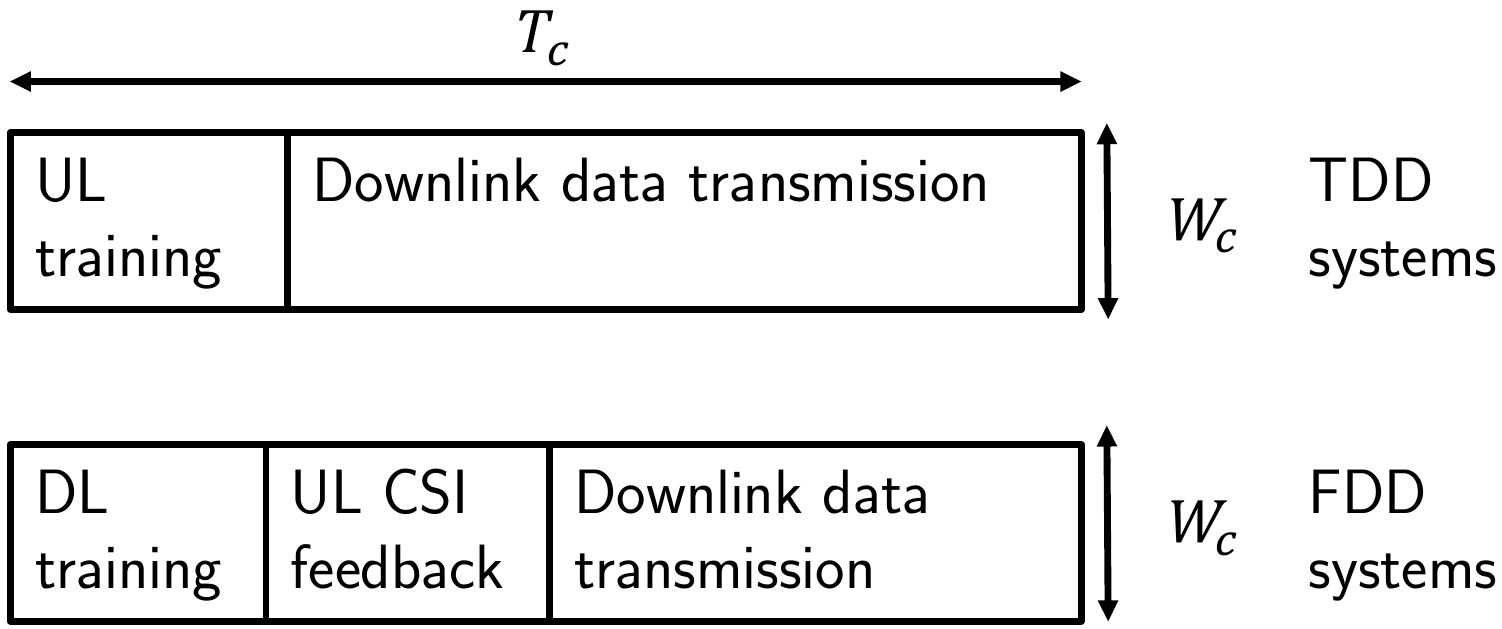}
\caption{Typical training and feedback procedures for TDD and FDD MU-MIMO systems.}
\label{Fig_training}
\end{figure}

CSIT is vital to harness the capacity gain in MU-MIMO systems. However, it comes at a price which is the overhead imposed by the CSIT acquisition process. CSIT is usually modeled as a matrix of channel coefficients in a narrow-band system, representing the base band complex channel gain between base station (BS) antennas and user-terminals. Without CSIT, the BS cannot separate signals for different user-terminals by distinct spatial modes, thus causing serious inter-user interference, which effectively eliminates the multiplexing gain of MU-MIMO systems. Moreover, CSIT has to be obtained in a timely and accurate manner, i.e., channel estimation has to be done for each user-terminal within its distinct channel coherence time, and the length of the training sequence should scale with the number of users\footnote{For time-division duplex (TDD) systems.}, or the number of transmit antennas\footnote{For frequency-division duplex (FDD) systems.}, respectively \cite{Hassibi03}. Fig. \ref{Fig_training} illustrates typical CSIT acquisition procedures in cellular systems. At the beginning of a transmission frame, the BS either listens to the uplink channel training sequences (for calibrated TDD systems), or first transmits downlink training sequences and then waits for CSIT feedback from users (for FDD systems or uncalibrated TDD systems). Then the BS starts downlink data transmission leveraging the obtained CSIT estimations. Such a pilot-assisted (or training-based) transmission scheme is widely adopted in MU-MIMO systems.

More often than not, it is conveniently assumed that CSIT overhead is negligible, and thus MU-MIMO systems can accommodate a large number of users, especially in massive MIMO systems, where a vast excess number of BS antennas are deployed and the TDD mode is adopted to exploit the channel reciprocity \cite{Marzetta10}. Even in this scenario, a significant dimensionality loss due to the CSIT acquisition overhead exists if the user channel coherence time is small, or the number of users in the cell is large. Such dimensionality loss calls for user-scheduling in MU-MIMO systems, i.e., user channels are only required on demand, which motivates our work.

In existing literature, substantial amount of work has been done studying the MU-MIMO downlink scheduling problem, most of which focuses on maximizing the network throughput given \emph{perfect CSIT} and \emph{CSIR} \cite{Huh11,Huh12,Pap10,Jang07,Raymond09}. In \cite{Raymond09,Jang07}, the MIMO downlink with multiuser scheduling is considered where users are equipped with multiple antennas, by which they perform receive-beamforming to counteract inter-user interference (ICI). In this way, the scheduler chooses users with larger channel magnitudes and better orthogonality. However, precoding is unnecessary in their setting, when the number of user antennas is sufficient to cancel the ICI, i.e., no CSIT is needed. In \cite{Pap10,Huh11,Huh12}, the MU-MIMO downlink scheduling with single-antenna users is considered where it is BS's responsibility to eliminate the ICI by precoding. The users are divided into groups based on their locations, and the scheduling is performed by selecting different user-groups. However, the acquisition of CSIT is incorporated into the scheduling decisions in none of the work above, where only \cite{Huh12} briefly discusses the impact of imperfect CSIT, whereas no scheduling schemes are given accordingly. Regarding CSIT acquisition overhead and imperfect CSIT, the MIMO system capacity under a general block-fading model without presuming a pilot-assisted scheme is considered in \cite{Marzetta99}. The seminal work \cite{Hassibi03} establishes the training capacity of a MIMO link and gives a lower bound of the capacity w.r.t. imperfect channel estimation, assuming the pilot-assisted scheme is adopted. Furthermore, the work \cite{Caire10} gives an achievable sum rate considering a practical training and feedback scheme with a zero-forcing precoder. To relieve the burden of CSIT estimation, the work \cite{Biguesh06} gives a comprehensive study on the training scheme design in the MIMO system. In \cite{Caire03}, the authors point out that without CSIT, the degree-of-freedom (DoF) of MU-MIMO systems falls off to $1$, which is identical with a single-antenna link. The user scheduling scheme with ``predictable'' and ``non-predictable'' CSI quality is considered in \cite{Shirani10}, where the Lyapunov technique \cite{Neely10} is first introduced to solve the user scheduling problem in MU-MIMO systems. Compared with the work above on MU-MIMO downlink scheduling, the current work is the first work to incorporate the CSIT acquisition overhead into the scheduling decisions, as far as we know.

In this paper, we investigate the design of user-scheduling schemes to maximize the throughput of MU-MIMO system downlink, on account of the user queue information (UQI) and the channel acquisition dimensionality loss. A unique issue that we address is that users have distinct channel coherence times\footnote{This is due to different user mobilities and scattering environments.}. The main contributions of this paper are:
\begin{itemize}
\item
Based on the Lyapunov-drift optimization, we formulate the generic user-scheduling problem as the genie-aided optimization problem (GAP). The corresponding user scheduling scheme, referred to as the GAP-rule, is proved to be throughput-optimal, i.e., it stabilizes the system as long as the arrival rates are inside the capacity region.
\item
Furthermore, two modified GAP-rule-based schemes are proposed. The $T$-frame dynamic channel acquisition (T-DCA) scheme is proposed to reduce the overhead related to the variable frame structure of the GAP-rule. A delay-fairness enhanced scheme, namely the power-law DCA scheme (PL-DCA) is also proposed. The throughput-optimality is proved for both schemes, respectively. Numerical results show significant improvements by both schemes dealing with corresponding concerns.
\item
In view of the fact that the GAP-rule-based schemes are practically infeasible due to their complexity and non-causality, we propose heuristic schemes, namely the queue-based quantized-block-length user scheduling scheme (QQS), $T$-frame QQS (T-QQS) and power-law QQS (PL-QQS), which substantially reduce the complexity and also make the GAP-based schemes practically feasible, respectively. It is shown by simulations that the QQS is asymptotically throughput-optimal under the conditions that the system dimension is large and the users are naturally grouped on account of their channel coherence times.
\item
We provide a new throughput-optimality proof for the high order polynomial Lyapunov function, which is especially useful for the PL-DCA scheme. Existing work uses the fluid-limit technique to prove the throughput-optimality. Our proof is based directly upon the Lyapunov-drift analysis, and thus is simpler to understand and gives more insights.
\end{itemize}

The paper is organized as follows. Section \ref{sec_SM} describes the system model and gives some preliminary knowledge. In Section \ref{sec_ME}, a motivating example is given to illustrate why we need to do DCA in MU-MIMO downlinks. In Section \ref{sec_MWT}, the GAP is formulated to maximize the Lyapunov-drift. Then the T-DCA and the PL-DCA are proposed. In Section \ref{sec_QQS}, we propose the QQS-based schemes. Section \ref{NR} gives the numerical results. Finally, in Section \ref{sec_c}, we draw the conclusions. Throughout the paper, we use boldface uppercase letters, boldface lowercase letters and lowercase letters to designate matrices, column vectors and scalars respectively.
\section{System Model and Preliminaries}
\label{sec_SM}
\begin{figure}[!t]
\centering
\includegraphics[width=0.4\textwidth]{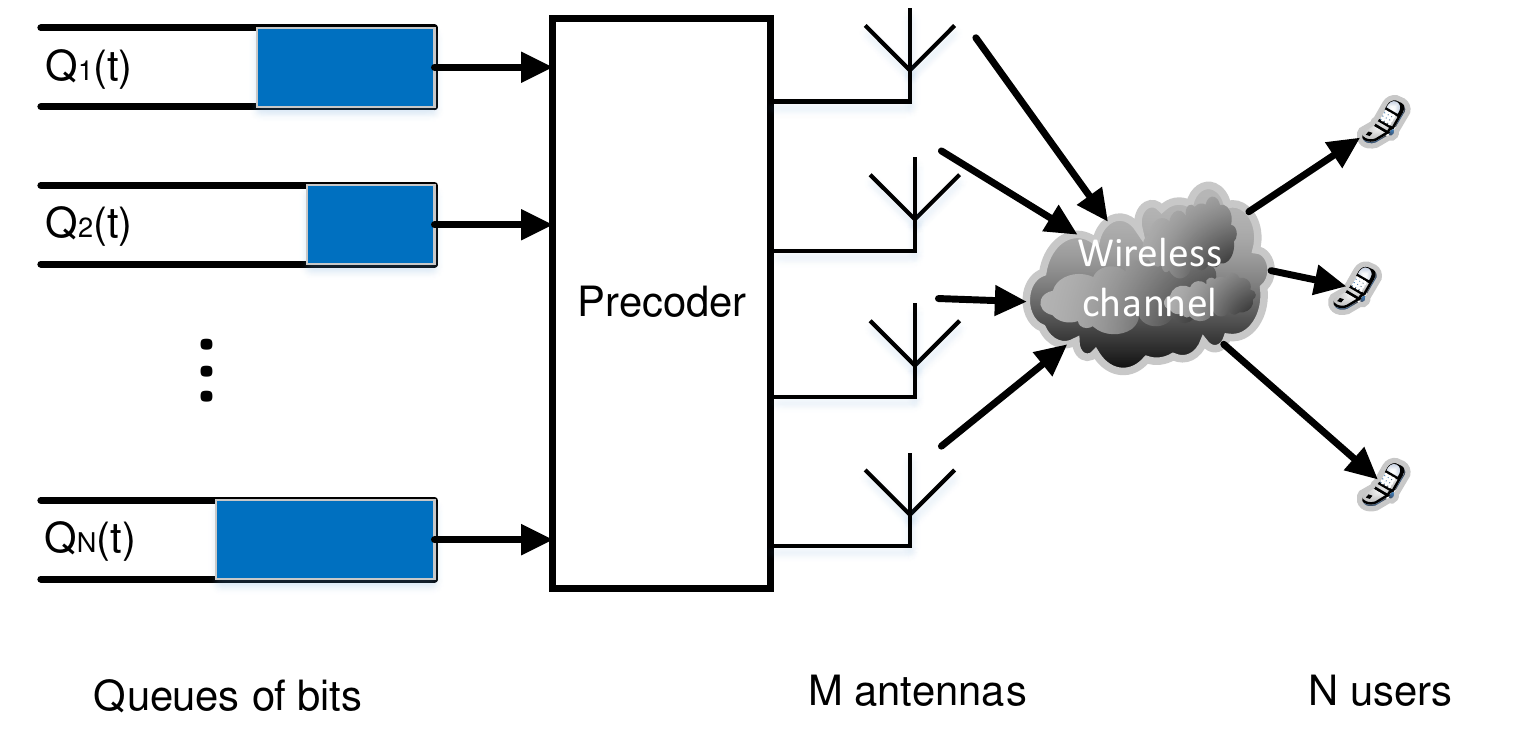}
\caption{System model of a MU-MIMO downlink scenario.}
\label{Fig_sysmodel}
\end{figure}
We consider the downlink (forward-link) of a single cell where an $M$-antenna BS serves $N$ single-antenna users, as shown in Fig. \ref{Fig_sysmodel}. We consider a narrow-band channel, and one channel use\footnote{A channel use, or a time slot, corresponds to an independent complex signal-space dimension in the time-frequency domain.} is characterized as
\begin{equation}
\label{channel_model_MUMIMO_vector}
\bm{y}(t)  =  \bm{H}(t)\bm{x}(t) + \bm{z}(t),
\end{equation}
where $\bm{x}(t) \in \mathbb{C}^M$ denotes the complex transmit signal vector of $M$ antennas at the BS, $t$ in the bracket denotes the index of channel use, $\bm{y}(t) \in \mathbb{C}^N$ denotes the receive signal vector of $N$ single-antenna users, $\bm{z}(t)$ denotes the cyclic symmetric zero mean complex Gaussian additive noise, i.e., $\bm{z}(t) \sim \mathcal{CN}(\bm{0},\sigma^2\bm{I}_N)$, and $\bm{H}(t) \in \mathbb{C}^{N \times M}$ denotes identically independently distributed (i.i.d.) Rayleigh fading coefficients with unit-norm entries. In particular, we consider linear precoding, where\footnote{Notice that $\bm{W}(t)$ can be any general linear precoding matrix, whereas we adopt the zero-forcing precoding matrix, i.e., $\bm{W}(t)=\bm{H}(t)^\dag(\bm{H}(t)\bm{H}(t)^\dag)^{-1}$ in the simulations.}
\begin{equation}
\label{LP}
\bm{x}(t) = \zeta(t) \bm{W}(t)\bm{s}(t),
\end{equation}
where $\zeta(t)$ is the power normalization factor with $\zeta(t)^2=\frac{P}{\textrm{tr}(\bm{W}(t)\bm{W}(t)^\dag)}$, $P$ is the total transmit power, $\bm{W}(t)$ is the precoding matrix, and $\bm{s}(t)$ denotes the i.i.d. user data streams. The signal-to-interference-noise ratio (SINR) of user-$n$ is written as,
\begin{equation}
\label{LP_sinr}
\gamma_n(t) = \frac{\zeta(t)^2 \left|\bm{h}_n^\dag(t)\bm{w}_n(t) \right|^2}{\sum_{j \neq n}{\zeta(t)^2 \left|\bm{h}_n^\dag(t)\bm{w}_j(t) \right|^2}+\bm{z}_n(t)^2},
\end{equation}
where we write $\bm{H}(t) = \left[\bm{h}_1(t),\bm{h}_2(t),...,\bm{h}_N(t)\right]^\dag$ and $\bm{W}(t)=\left[\bm{w}_1(t),\bm{w}_2(t),...,\bm{w}_N(t)\right]$, and $n$ is the user index. Furthermore, let $Q_n(t)$ denote the queue length in bits of user $n$ at the beginning of $t$-th channel use, let $a_n(t)$ denote the number of arrival bits from upper layer to the physical layer between $(t-1)$-th and $t$-th channel uses, and let $\mu_n(t)$ denote the allocated number of service bits to Queue-$n$, which equals the allocated service rate (bits/channel use) in this case. Then the queuing dynamics are written as
\begin{equation}
\label{QueueD}
{Q_n}(t+1) = {Q_n}(t) - {\tilde \mu _n}(t) + {a_n}(t),
\end{equation}
where ${{\tilde \mu }_n}(t) = \min \{ {Q_n}(t),{\mu _n}(t)\}$ denotes the actual service number of bits, considering the circumstances that sometimes the queue is emptied given the amount of allocated service bits.
\newtheorem{defn}{Definition}
\begin{defn}
Queue-$n$ is said to be strongly stable if \cite{Neely10}
\begin{equation}
\label{QueueS}
\mathop {\limsup }\limits_{T \to \infty } \frac{1}{T}\sum\limits_{t = 1}^T \mathbb{E} [{Q_n}(t)] < \infty,
\end{equation}
when there is no bound on the buffer size for any $n$.
\end{defn}

When all queues are strongly stable in the system, the time-average actual service rate equals the arrival rate, i.e.,
\begin{equation}
\label{QueueR}
\mathop {\lim }\limits_{T \to \infty } \frac{1}{T}\sum\limits_{t = 1}^T {{{\tilde \mu }_n}(t)}  = \mathop {\lim }\limits_{T \to \infty } \frac{1}{T}\sum\limits_{t = 1}^T {{a_n}(t)}, \, \forall n.
\end{equation}
Notice that the left-right-side is the time average of the realizations of the actual service rate, thus we do not need the expectation to hold (\ref{QueueR}).

The achievable ergodic rate region $\mathcal{R}$ is defined as the convex hull of all achievable rate points of $n$ users. Denote all feasible transmission schemes as $\mathcal{X}$ and the transmission scheme $\pi_s \in \mathcal{X}$, where $s$ is the index for scheduling policies, is the user scheduling scheme and the corresponding precoding scheme with the rate of user-$n$ at time t,
\begin{equation}
{R_n}(\bm{H}(t),{\pi _s}(t))=\bm{I}({\pi _s}(t))\log(1+\textrm{SINR}(\bm{H}(t),\pi_s(t))),
\end{equation}
where $\bm{I}({\pi _s}(t))$ is an indicator function which determines whether user-$n$ is scheduled, and $\textrm{SINR}(\bm{H}(t))$ is the signal to noise and interference ratio which is related to the channel realization and the precoding scheme.\footnote{Explicit expressions of ${R_n}(\bm{H}(t),{\pi _s}(t))$ will be shown later in Section \ref{sec_MWT}} The user-$n$ achievable rate is defined as the time-average of user rate
\begin{equation}
\label{rateR}
{\bar R_n} = \mathop {\lim }\limits_{T \to \infty } \frac{1}{T}\sum\limits_{t = 1}^T {{R_n}(\bm{H}(t),{\pi _s}(t))} ,\, \forall n.
\end{equation}
Based on ergodicity, \eqref{rateR} equals
\begin{equation}
\label{Egrate}
{{\bar R}_n} = \mathbb{E}\{ {R_n}(\bm{H},{\pi _s})\} ,\, \forall n,
\end{equation}
where the expectation is taken over all possible channel gain $\bm{H}(t)$ and possibly ${\pi _s}(t)$ when a randomized control policy is considered. The achievable ergodic rate region can be characterized as
\begin{equation}
\label{EgrateRegion}
\mathcal{R }= \textrm{coh}\bigcup\limits_{{\pi _s} \in \mathcal{X}} {\{ \bm{\bar R}:0 \le {{\bar R}_n} \le \mathbb{E}[{R_n}(\bm{H},{\pi _s})]} \},
\end{equation}
where $\bm{\bar R}$ is a $N$-dimensional region, ${{\bar R}_n} $ is its $n$-th component, and ``coh'' denotes the closure of a convex hull.
\begin{defn}
\label{to}
(\emph{Throughput-Optimality}) A scheduling scheme is throughput-optimal if for any arrival rate point inside the achievable ergodic rate region, the system can be stabilized by the scheduling scheme.
\end{defn}

Note that the throughput in this paper refers to the downlink throughput, not concerning uplink throughput. We consider a generic scenario where each user has its distinct block length which denotes the number of consecutive channel uses that the user-channel stays static, or also referred to as channel coherence time. The block fading channel model is adopted in this paper, where every user's channel stays constant for $T_n$ consecutive channel uses, and changes to another constant according to an i.i.d. (over time and users) random process. Denote by $T_n$ the channel coherence time of user-$n$, and let $\mathcal{T}=\{T_1,T_2,...,T_N\}$\footnote{The distinction of user block lengths is due to the fact that there are several factors that can affect the block length of each user, such as distinct user-mobility, scattering environment nearby, frequency offset and etc. We assume the BS knows the channel coherence time \emph{a priori}, since channel coherence time is a second-order channel statistics, which can be regarded to be static for a relatively long period.}. Notice that there are no units for both $T_c$ and $T_k,\forall k$, since by definition, the block length in block fading model equals channel coherence time multiplied by channel coherence bandwidth. Therefore the block length describes the channel coherence, both temporally and spectrally. Similar notations have been used in \cite{Kobayashi11}\cite{Hassibi03}.

Here we assume that for every $T_n$ channel uses, the system adopts one uplink channel training symbol (reciprocal channel is considered here) to estimate the channel of user-$n$. By doing this, we assume the BS obtains the \emph{perfect CSIT} of user-$n$. Hence, the number of concurrent scheduled users, denoted by $N_\textrm{s}$, equals the number of channel uses for channel estimation numerically in each fading block. Note that this is actually an optimistic assumption on the MU-MIMO system, since normally we can only get a noisy version of the CSIT, and the system has to do the channel training more than once to refine the estimation. Nonetheless our results can be extended to this scenario immediately by multiplying the training length with a predefined factor, taking into account the imperfection of channel estimation \cite{Caire10}.

In the following section, a simple example is given to illustrate why we need user scheduling in MU-MIMO systems, in contrast to most existing literature, which assumes simultaneous transmission to all users especially in the massive MIMO system. The impact of limited block length is considered. Qualitatively speaking, when $T_n$ is very small, it would be even more advantageous to leverage a space-time coding (STC) scheme \cite{Tse05} to serve one user at a time without CSIT, alleviating the prohibitively large cost of channel acquisition.
\begin{figure}[!t]
\centering
\includegraphics[width=0.4\textwidth]{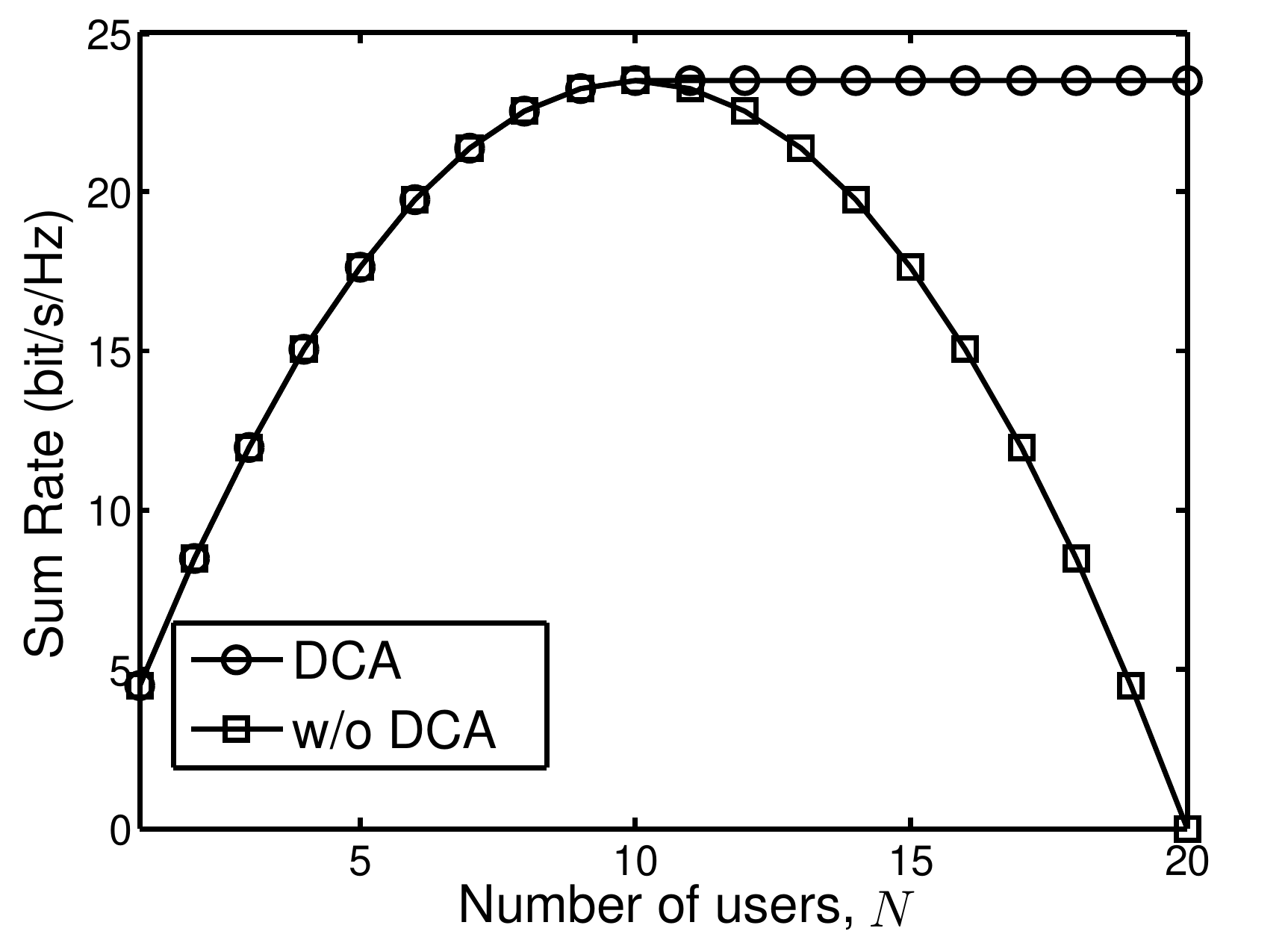}
\caption{Cell throughput with and without DCA. We assume each active user occupies one channel use to transmit the uplink training symbol. Therefore, the CSI acquisition overhead equals to the number of users, and $T_n=20,\,\forall n$.}
\label{Fig_TI}
\end{figure}
\section{Motivations}
\label{sec_ME}
In principle, user-scheduling in MU-MIMO is of great necessity when the total channel acquisition overhead is comparable to the channel coherence time. From Fig. \ref{Fig_TI}, where we assume $M \rightarrow \infty$, we can observe that, clearly, simultaneous transmission to all users is detrimental when CSIT acquisition overhead is large. The DCA scheme used in Fig. \ref{Fig_TI} is that at each scheduling step, randomly schedule $10$ users when the number of users is larger than $10$. This scheme is DoF-optimal by the following arguments. The DoF of the downlink BC is $\min[M,N_s]$, and the training DoF, which is defined as the DoF considering the CSIT acquisition overhead, is the DoF of BC multiplied by a factor, $\frac{T_\textrm{c}-N_\textrm{s}}{T_\textrm{c}}$, assuming all users have identical channel coherence time $T_\textrm{c}$. Therefore, the training DoF is, in the massive MIMO regime,
\begin{equation}
\label{DoF_t}
\frac{T_\textrm{c}-N_\textrm{s}}{T_\textrm{c}} \min[M,N_\textrm{s}] = \frac{T_\textrm{c}-N_\textrm{s}}{T_\textrm{c}} N_\textrm{s} \le \frac{T_\textrm{c}}{4},
\end{equation}
where $N_\textrm{s}$ is the number of concurrent active users. The equality in \eqref{DoF_t} holds when $N_\textrm{s} = \frac{T_\textrm{c}}{2}$. Therefore, simultaneous transmission to $10$ users, which equals the number of half the channel coherence time, is DoF-optimal.

Furthermore, the following intuition is true, that simultaneous transmission to users with dramatic channel coherence time difference is not desirable. To illustrate, suppose there are $40$ users in the system, with channel coherence times
\begin{equation}
\label{CohrTime}
T_1=...=T_{39}=50\textrm{, and } T_{40}=5.
\end{equation}
Assuming the user rates (when scheduled) all equal to $1$, not affected by channel-fading, which is the case when $M$ is large due to the channel hardening effect \cite{Marzetta10}. Consider a time sharing DCA scheme, where the general sum rate can be formulated as
\begin{equation}
\label{GF_TS}
\textrm{Sum Rate}  = \sum_{i=1}^D \sum_{n \in \mathcal{U}_i} p_i \left(1 - \mathcal{I}(|\mathcal{U}_i| > 1)\sum_{n \in \mathcal{M}_i} \frac{1}{T_n} \right),
\end{equation}
where the total number of time-sharing transmission modes is $D$, denote $\mathcal{U}_i$ as the set of users which are spatial multiplexed by the MU-MIMO transmission in the $i$-th mode, denote $p_i$ as the percentage of time allocated to mode $i$, with $\sum_i{p_i}=1$, denote $|\cdot|$ as the cardinality of a set, and $\mathcal{I}(\cdot)$ is the indicator function. Note that when $|\mathcal{U}_i| = 1$, the BS adopts the STC scheme to serve the only user, where CSIT is not required, and that for each spatial multiplexing transmission mode with $|\mathcal{U}_i| > 1$, the time-frequency resources dedicated to channel estimation is $\sum_{n \in \mathcal{U}_i} \frac{1}{T_n}$ since the CSIT has to be estimated every $T_n$ channel uses for user $n$. Based on \eqref{GF_TS}, the no-DCA scheme, i.e., spatial multiplexing all users renders
\begin{equation}
\label{TP}
\textrm{Sum Rate}  = (1 - \frac{1}{5} - 39\times\frac{1}{{50}} )\times 40 = 0.8.
\end{equation}

On the other hand, consider a time-sharing DCA scheme, where there are two modes. One is to transmit to user $(1,2,...,39)$ with $p_1=80\%$, while the other is to transmit to user $40$ exclusively with the STC mode and $p_2=20\%$, whereby,
\begin{equation}
\label{TSrate}
\textrm{Sum Rate}  = 0.8 \times (1-39 \times \frac{1}{50}) \times 39 + 0.2 = 7.064,
\end{equation}
which is approximately $10$-fold compared with the no-DCA scheme. Note that this time-sharing scheduling scheme does not need the UQI because the arrival rates are known to the BS. However, in practice, the user arrival rates cannot be known \emph{a priori}. Under this circumstance, the UQI is leveraged to facilitate the DCA, which is discussed in the following sections.
\section{Genie-Aided Dynamic Channel Acquisition}
\label{sec_MWT}
In this section, we first formulate the generic DCA optimization problem, the GAP, which maximizes the Lyapunov-drift every scheduling step with the aid of a genie who provides the BS the instantaneous channel coefficients before channel estimation. The resulting scheduling scheme, albeit practically infeasible, is termed as the \emph{GAP-rule}. Then we provide the throughput-optimality proof for the GAP-rule and propose two other scheduling schemes dealing with implementation concerns, which are both proved to be throughput-optimal. Due to the throughput-optimality of the GAP-rule based schemes, they serve as performance bounds in this paper. In the next section, we will propose heuristic algorithms, which are practically feasible versions of the aforementioned GAP-rule based schemes and also show near-optimal performance. Note that we assume the coherence times of all users are known to the BS, since the channel coherence time is usually changing slowly, about seconds to tens of seconds, and thus it can be estimated efficiently. In practice, the channel coherence time can be obtained from user mobility estimation, which is available in Long-Time-Evolution (LTE) systems \cite[Section 5.2.4.3]{3GPP09}.
\subsection{GAP}
The generic DCA optimization problem, namely GAP, is formulated based on the framework of \cite{Neely10}. To stabilize the system whenever the arrival rate is inside the achievable rate region, the optimization boils down to select the users which optimize the Lyapunov drift in each scheduling step, i.e.,
\begin{IEEEeqnarray}{rll}
\label{MWTR}
\mathop {\textrm{maximize}}\limits_{\mathcal{S} \subseteq \mathcal{N}} \quad\quad & \left[\sum\limits_{n \in \mathcal{S}} {\frac{{Q_n({t_k}){\beta _n}({t_k})}}{{T_k}}}\right],
\end{IEEEeqnarray}
where $\mathcal{S}$ is the optimization variable which denotes the set of scheduled users, $\mathcal{N}$ is the overall user set.
\begin{equation}
{\beta _n}({t_k})=  \left\{ \,
\begin{IEEEeqnarraybox}[][c]{l?s}
\IEEEstrut
(T_k-|\mathcal{S}|) r_n^{\textrm{SM}}(t_k ) & if $|\mathcal{S}|>1$, \\
T_k r_n^{\textrm{STC}}(t_k ) & if $|\mathcal{S}|=1$,
\IEEEstrut
\end{IEEEeqnarraybox}
\right.
\label{ADB}
\end{equation}
where ${\beta _n}({t_k})$ denotes the allocated service bits of user $n$ at time $t_k$, and
\begin{IEEEeqnarray}{rl}
\label{rate_term1}
r_n^{\textrm{SM}}(t_k ) &= \log\left(1+\gamma_n^{(t_k)}\right),\\
\label{rate_term2}
{r^{\textrm{STC}}_n}(t_k) &= \log \left(1 + \frac{{{{\left\| {{\bm{h}_n{(t_k)}}} \right\|}^2}P}}{{M{\sigma ^2}}}\right),
\end{IEEEeqnarray}
and
\begin{equation}
T_k=  \left\{ \,
\begin{IEEEeqnarraybox}[][c]{l?s}
\IEEEstrut
\mathop{\min}\limits_{n \in \mathcal{S}} \left[{T_n}\right] , & if $|\mathcal{S}|>1$, \\
T_{\textrm{STC}}, & if $|\mathcal{S}|=1$.
\IEEEstrut
\end{IEEEeqnarraybox}
\right.
\label{framelength}
\end{equation}
The objective in \eqref{MWTR} can be seen as the queue-size-weighted sum of the user service rates. Since we assume the users each occupies one uplink training channel use to obtain a perfect CSIT, the amount of time-frequency resources dedicated to channel estimation is the number of concurrent users and the remainder is $T_k - |S|$ in \eqref{ADB}. $T_{\textrm{STC}}$ is a predefined constant. When the number of selected users is larger than one, spatial multiplexing is enabled with user rate $r_n^{\textrm{SM}}(t_k)$ and channel estimation overhead $|\mathcal{S}|$. Otherwise, STC is leveraged to serve one user at a time with rate ${r^{\textrm{STC}}_n}(t_k)$.

\begin{figure}[!t]
\centering
\includegraphics[width=0.5\textwidth]{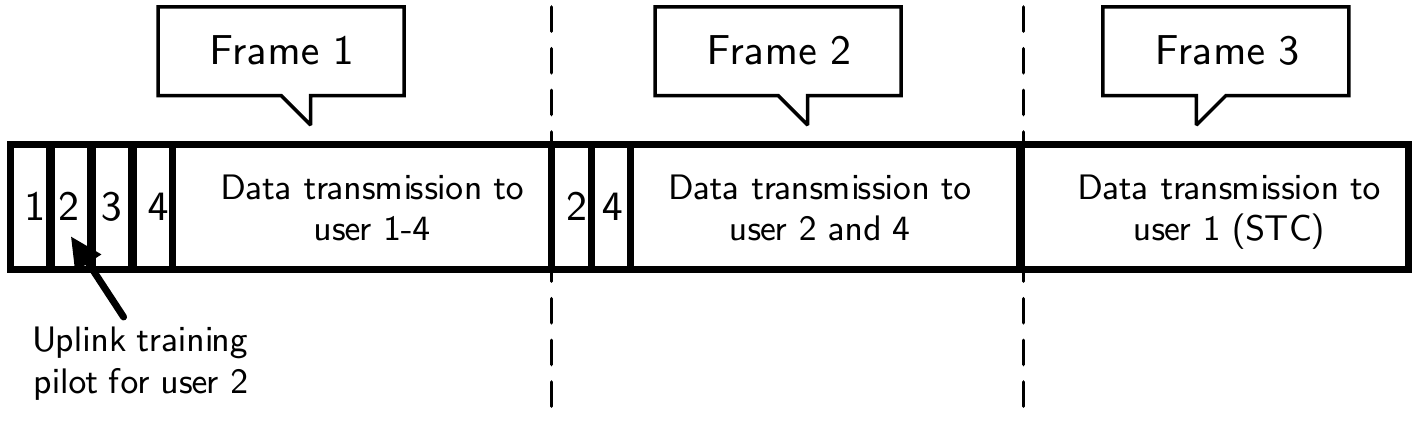}
\caption{A sample path of the control scheme. Only three frame transmission is shown for simplicity.}
\label{Fig_SP}
\end{figure}

Notice that the GAP-rule and several schemes introduced in the following adopt a variable frame-length design. A sample path of the scheduling scheme is shown in Fig. \ref{Fig_SP}. If multiple users are chosen, the chosen users have to send training pilots first to let the BS have the CSIT. The frame length when multiple users are chosen is set to be the \emph{minimum channel coherence time} of the selected users. In this way, during one frame, the channel estimations of all scheduled users are meaningful. On the other hand, if only one user is chosen, the BS will use the STC scheme, with \emph{no channel estimation} needed. Note that the assumption of the frame length when multiple users are selected actually makes the resulting user rates a lower bound of the system capacity since some users may have remaining channel coherence times and thus do not need to do channel estimation immediately. However, this assumption makes the Lyapunov drift analysis tractable, in the sense that otherwise, the decisions of different scheduling steps would be dependent due to the possible remaining channel coherence times of some users, which makes the analysis much more difficult.

Specifically, the frame-by-frame queueing dynamics are written as
\begin{equation}
\label{QD}
{Q_n}({t_{k + 1}}) = \max[{Q_n}({t_k}) - {\beta _n}({t_k}),0] + {\alpha _n}({t_k}),\,\forall n,
\end{equation}
where $t_k$ denotes the beginning of frame-$k$, ${\beta _n}({t_k})$ and ${\alpha _n}({t_k})$ denote the allocated service bits and arrival bits during the time interval $[t_k,t_{k+1})$, respectively.

The infeasibility of the GAP-rule can be specified as follows. It is clear that the rates in \eqref{rate_term1} and \eqref{rate_term2} \emph{cannot} be evaluated to proceed the optimization in practice unless we have a genie who provides the BS all the channel coefficients without having to do the channel estimation. Most existing literature assumes the CSIT is known \emph{a priori} \cite{Yoo06} without considering the acquisition overhead, or coarse knowledge of CSI is available \cite{Zakhour07}, while neither of which is practical considering the CSIT overhead. Even supposing the genie is available, the GAP-rule is still NP-hard under generic linear precoding, observing that we have to exhaustively search all the user sets to obtain the optimum. Therefore, in the next section, we will propose heuristic algorithms, which are practical with much less complexity, and meanwhile present little to none performance degradation. In what follows, we will prove the throughput-optimality of the GAP-rule.

\newtheorem{theorem}{Theorem}
\begin{theorem}
\label{Thm_TO}
(\emph{Throughput-Optimality of the GAP-Rule}) Suppose $a_n(t_k)$ is i.i.d. over time and satisfies
\begin{equation}
\label{amax}
0 \le a_n(t_k) \le A_{\textrm{max}},\,\forall n, k
\end{equation}
under the frame design described in Fig. \ref{Fig_SP}, the GAP-rule is throughput-optimal.
\end{theorem}
\begin{IEEEproof}
The proof is based on the framework of \cite{Neely10}, with the difference that our scheme adopts a variable frame-length structure. We need to specify that the proof is still effective in this circumstance. The details are given in Appendix \ref{proof_thm1}.
\end{IEEEproof}
\begin{corollary}
\label{Thm_TR}
(\emph{The Ergodic Sum Capacity of the GAP-Rule}) The ergodic sum capacity can be computed by running the following admission control and scheduling schemes:\\
\textbf{Admission control}: Before each frame, for every queue with queue size $Q_n(t_k)<V$, the number of arrival bits is set to be $W_{\textrm{max}}$, where $V$ and $W_{\textrm{max}}$ are constants\footnote{For Typical values, $V$ and $W_{\textrm{max}}$ can be approximately $100$-fold of the arrival rate.}. Otherwise, there are no arrival bits during this frame.\\
\textbf{Scheduling}: Schedule the users with the GAP-rule.\\
Then calculate the time-average sum arrival rate $A_{\textrm{avg}}$. We have
\begin{equation}
\label{SRO}
A_{\textrm{avg}} \ge R^*-\frac{B}{V},
\end{equation}
and the system is stable, where $B$ is a finite constant, and $R^*$ is the maximum ergodic sum rate, i.e., the ergodic sum capacity.
\end{corollary}
\begin{remark}
Leveraging Corollary \ref{Thm_TR}, by letting $V$ be sufficiently large, we can find the maximum ergodic sum rate of the GAP-rule. This result can be straightforwardly generalized to other scheduling schemes. Theorem \ref{Thm_TO} establishes a throughput-optimal scheduling scheme, whereas it is still unknown how to characterize the achievable throughput explicitly. To this end, we provide Corollary \ref{Thm_TR} to calculate the maximum ergodic sum rate, which will help us demonstrate the performance of proposed schemes in Section \ref{NR}.
\end{remark}
\subsection{$T$-Frame DCA Scheme (T-DCA)}
The scheduling frequency should be kept reasonably low in practice, on account of the complexity issue and the additional signalling overhead due to variable frame-length. In this regard, we propose the T-DCA, based on which the users are scheduled, still according to the GAP-rule, however only every $T$ frames, where $T$ is a predefined positive integer. In this way, the scheduling complexity and signalling overhead scales down linearly with $T$. Meanwhile, the T-DCA is still throughput-optimal, which is shown by the following corollary.
\begin{corollary}
\label{coro1}
(\emph{Throughput-Optimality of T-DCA}) Under the exact same conditions defined in Theorem \ref{Thm_TO}, the T-DCA scheme, which schedules the users according to the GAP-rule every $T$ frames, is throughput-optimal.
\end{corollary}
\begin{IEEEproof}
The key notion is that the Lyapunov-drift of this T-DCA scheme is within a constant to the optimal control scheme, and based on the \emph{C-addictive approximations} technique developed in \cite{Neely10}, it can be proven that the scheme is throughput-optimal. See Appendix \ref{proof_thm3} for details.
\end{IEEEproof}
\begin{remark}
Despite the fact that the T-DCA scheme can reduce the implementation overhead, the user-delay will increase due to the ``lazy'' scheduling policy of the T-DCA scheme. By carefully designing the parameter $T$ in the T-DCA scheme, we can strike a good balance between user average delay performance and the signaling overhead imposed by frame-structure modifications. The numerical results will be shown later in Section \ref{NR}.
\end{remark}
\begin{remark}
One drawback we observe during running GAP-rule or T-DCA is that generally they encounter user \emph{delay-unfairness}. Notice that these schemes can only guarantee that the total time-average expected queue size is finite \eqref{QueueStk}, the delay-fairness among users is not guaranteed. In fact, the per-user average-delay profile varies dramatically with each other by these schemes. The reason can be roughly explained by the following example. Suppose there are $4$ users in the system, $3$ of them have long coherence time, while the other has short coherence time. Consider the GAP-rule, which chooses the set of users that maximize the queue-size-weighted sum of service rates.The scheme prefers to serve the \emph{long-coherence-time (LT)} users simultaneously. While only when the queue of the \emph{short-coherence-time (ST)} user accumulates to be considerably large (approximately $3$ times larger than the others), the GAP-rule will schedule him/her, because it is not worth including the ST user into the spatial multiplexing (SM) mode since he/she will ``drain up'' the system resources by frequent channel estimation. Therefore, the average delay of the ST user will be much larger than the LT users. This delay-fairness can be improved without harming the delay performance of the LT users or the throughput-optimality of the schemes, observing that the queues of the LT users are actually ``over-served''. This is because although the GAP-rule manages to maximize the queue-size-weighted sum rate every frame, the queues of the chosen users are likely to be emptied during the frame, rendering the allocated service rates larger than the actual user-received service rates. This is referred to as ``over-service'' since some resources are wasted. To avoid this and to utilize these resources to serve the starving queues, we next propose a delay-fairness improved scheme.
\end{remark}
\subsection{The Power-Law DCA Scheme (PL-DCA)}
\label{sec_DF}
In this subsection, the PL-DCA scheme is proposed based on the observation that longer queues should get better chance to be served in this scenario, while maintaining the throughput-optimality.

The intuition of the PL-DCA scheme is to give the long queue a larger weight than that in the GAP-rule. Hence, the average queue sizes of the ST users will be smaller. According to the \emph{Little's law} \cite{Garcia08}, the average delay will be smaller accordingly. The PL-DCA scheduling rule is, in contrast with the GAP-rule,
\begin{equation}
\label{PLMW}
\textrm{maximize: } \sum\limits_n {\frac{{Q_n^\theta ({t_k}){\beta _n}({t_k})}}{{{T_k}}}} ,
\end{equation}
where $\theta > 1$.
\begin{theorem}
\label{Thm_PLMW}
(\emph{Throughput-Optimality of the PL-DCA Scheme}) For an odd integer $\theta$, and any arrival rate point inside the achievable ergodic rate region, the system under PL-DCA scheme is \emph{mean-rate stable}, i.e.,
\begin{equation}
\label{MRS}
\mathop {\lim }\limits_{t_k \to \infty } \sum\limits_n {\frac{{\mathbb{E}[{Q_n}(t_k)]}}{t_k} = 0},\,\forall n.
\end{equation}
\end{theorem}
\begin{IEEEproof}
See Appendix \ref{proof_thm2}.
\end{IEEEproof}
\begin{remark}
For the throughput-optimality of the scheduling scheme with power-law queue terms, the authors in \cite{Andrews04} give a proof for a slot-by-slot system leveraging a \emph{fluid-limit technique}. Our proof is applicable to the frame-by-frame control in this paper, and bases directly upon the \emph{Lyapunov-drift}. Only the case when $\theta$ is an odd integer is proved, whereas it is sufficient for the PL-DCA scheme to work since $\theta$ is only needed to be larger than one to achieve better delay-fairness.
\end{remark}
\begin{remark}
Remark that the throughput-optimality for all three aforementioned schemes proved in this paper is only applicable to the frame design described in Fig. \ref{Fig_SP}. Nonetheless, our proposed scheme under this frame design displays evident performance gain over the non-DCA scheme, as will be shown in Section \ref{NR}. One can do better if a slot-by-slot design is adopted, meaning the scheduling makes a decision upon every channel use. Nonetheless, this frame design has the advantage of making the control decision \emph{independent} over time, rendering the optimality proof tractable. To illustrate, considering the slot-by-slot design, one cannot make the scheduling decision independently because whether we can transmit to some set of users simultaneously depends on the validity of their CSIT, and thus depends on the historical scheduling actions.
\end{remark}
\section{Queue-Based Quantized-Block-Length Scheduling Scheme (QQS)}
\label{sec_QQS}
Due to the fact that the GAP-rule-based schemes described above, namely the GAP-rule, T-DCA, PL-DCA, require genie-aided CSIT, and they are NP-hard, the schemes are practically infeasible. To this end, we propose the QQS-based schemes, namely QQS, T-QQS, PL-QQS, corresponding to the practical versions of GAP-rule, T-DCA, PL-DCA, respectively. In this section, we will first specify the QQS, which bases its scheduling decision solely upon the UQI, neglecting the channel fading fluctuations. In addition, to reduce the complexity, we divide the users into groups according to their respective channel coherence times, and schedule among different groups, based on the intuition that serving users with significant channel coherence time difference is undesirable since the users with longer coherence time will be encumbered, as illustrated in Section \ref{sec_ME}. For T-QQS and PL-QQS, exact same techniques are used to make the T-DCA and PL-DCA practical, and corresponding changes are specified analogous to the QQS. Note that we neglect the time index in the following algorithm description. The QQS, which corresponds to the GAP-rule, is specified as
\begin{itemize}
\item
Step 1) Initialization:

Denote the overall user set by $\mathcal{N}$. Divide the users into $K$ groups, each of which denoted by $\mathcal{N}_k$, $k=1,2,...,K$, based on a uniform channel coherence time quantization
\begin{equation}
\label{CHTQ}
\mathcal{N}_k = \left\{n \in \mathcal{N} \left| \frac{k-1}{K} T_{\textrm{max}} \le T_n \le \frac{k}{K} T_{\textrm{max}}\right.\right\},
\end{equation}
where $T_{\textrm{max}} = \max{[T_n]},\,\forall n$, and the users are indexed by
\begin{equation}
\mathcal{N}_k = \left\{k_1,k_2,...,k_{|\mathcal{N}_k|}\right\},
\end{equation}
such that $Q_{k_1} \ge Q_{k_2} \ge...\ge Q_{k_{|\mathcal{N}_k|}}$. And
\begin{IEEEeqnarray}{rcl}
\label{Tbar}
\bar T_k &=& \mathcal{M}[T_n, n \in \mathcal{N}_k], \\
\mathcal{F}_k &=& \{k_1\},\\
i &=& 1,
\end{IEEEeqnarray}
where $\mathcal{M}(\cdot)$ denotes the empirical mean.
\item
Step 2) Group Selection: \\
For $k=1:K$,

\hspace{8mm} For $i=1:|\mathcal{N}_k|$,

\hspace{8mm} If
\begin{equation}
\label{iteration}
\left(1-\frac{i+1}{\bar T_k}\right)Q_{k_{i+1}}-\frac{1}{\bar T_k} \sum_{n=1}^i Q_{k_n}>0,
\end{equation}
\hspace{8mm} let
\begin{IEEEeqnarray}{rcl}
\mathcal{F}_k &=& \mathcal{F}_k \cup \{k_{i+1}\},\\
i &=& i+1,
\end{IEEEeqnarray}
\hspace{8mm} Else, break for.

\hspace{8mm} End for.

End for.
\item
Step 3): For each group $k$, compute
\begin{IEEEeqnarray}{rcl}
\label{pk}
\mathcal{P}_k &=& \max\left[ \left\{\left(1-\frac{|\mathcal{F}_k|}{\bar T_k}\right) \sum_{n\in \mathcal{F}_k} Q_n\right\}\right. \nonumber\\
&&\left.\bigcup \left\{Q_{j},\,\forall j \in \mathcal{N}_k\right\}\right],
\end{IEEEeqnarray}
and set
\begin{equation}
\label{fk}
\mathcal{F}_k = \{k_j\},
\end{equation}
only if the maximization in \eqref{pk} finds its maximum at a single queue length, $Q_{k_j}$.\\
Let
\begin{equation}
k^*=\textrm{argmax}[\mathcal{P}_k].
\end{equation}
\item
Step 4): Output the scheduled user set $\mathcal{F}_{k^*}$.
\begin{flushright}\rule{3mm}{3mm} \end{flushright}
\end{itemize}

Several technical details of the QQS should be mentioned. The reasoning for choosing \eqref{iteration} is that since we assume the channel coherence times for users in the same group are approximately identical \eqref{Tbar} and we neglect channel fluctuations. We asses whether it is worth adding the $(i+1)$-th user in group $k$, i.e., we compare the Lyapunov-drift of scheduling user $\{k_1,...,k_{i+1}\}$ with scheduling user $\{k_1,...,k_i\}$,
\begin{IEEEeqnarray}{rcl}
\label{R31}
\mathcal{L}(i+1)-\mathcal{L}(i) &=& \sum_{n=1}^{i+1} \left(1-\frac{i+1}{\bar T_k}\right)Q_{k_{n}}r_{k_{n}}^{\textrm{SM}}\nonumber\\
&-&\sum_{n=1}^{i} \left(1-\frac{i}{\bar T_k}\right)Q_{k_{n}}r_{k_{n}}^{\textrm{SM}}.
\end{IEEEeqnarray}
It is observed that whether \eqref{R31} is positive or negative is irrelevant with $r_{k_{n}}^{\textrm{SM}}$ because we assume $r_{k_{n}}^{\textrm{SM}},\forall n,k$ are identical by design of the QQS. Therefore, \eqref{iteration}  stems from \eqref{R31} immediately. For \eqref{pk}, $\mathcal{P}_k$ denotes the pre-log factor of the queue-weighted sum rate for the $k$-th group after we select the scheduled set $\mathcal{F}_k$, considering the possibility that scheduling one user with STC mode is the better choice, which results in the union with $Q_{k_i}$ in \eqref{pk}. By selecting the maximal $k^*=\textrm{argmax}[\mathcal{P}_k]$, we find the optimal scheduled group of users, within the heuristics of the algorithm.
\begin{remark}
It is clear that the computational complexity of the QQS algorithm is $\mathcal{O}(N)$ because it only involves running a sequential test of all users. The GAP-based algorithms are $\mathcal{O}(2^{N})$ because an exhaustive search over all user sets is involved. The reason for the dramatic complexity decrease compared with the GAP-rule is two-fold. First we group the users based on their channel coherence times, and treating the channel coherence times of users in each group as identical. Note that in practice, such a grouping is reasonable since users are usually categorized into several \emph{mobility states}, see e.g. \cite[Section 5.2.4.3]{3GPP09} for standardizations in the LTE system.

Secondly, we neglect the impact of channel fluctuations. Nevertheless, it can be anticipated that when the number of BS antennas becomes large, i.e., in massive MIMO systems, the user rates are no longer affected by small-scale channel fading, which is the so-called \emph{channel hardening effect} \cite{Marzetta10}. Therefore, the QQS is expected to be \emph{asymptotically throughput-optimal} in the large system regime. The effect will be shown in numerical results in Section \ref{NR}.
\end{remark}
\subsection{T-QQS and PL-QQS}
The QQS-based schemes for T-DCA and PL-DCA are termed as T-QQS and PL-QQS, respectively. The specifications for the T-QQS and PL-QQS are omitted for brevity, whereas it is straightforward to generalize from the QQS. Note that the T-QQS is scheduling the users according to the QQS every $T$ frame, and for the PL-QQS, replace all the queue terms, i.e., $Q_n(t)$'s $\forall \,n,t$, in the QQS with $Q_n^\theta(t)$'s, respectively.
\begin{table}[!t]
\label{APPpara}
\renewcommand{\arraystretch}{1.3}
\caption{system parameters}
\label{tab_para}
\centering
\begin{tabular}{ l r}
\hline
\bfseries Parameters & \bfseries Value \\
Carrier frequency $f_c$ & $2.6$~GHz  \\
Cell radius & $1000$~m \\
Bandwidth & $15$~KHz\\
Downlink SNR & $15$~dB\\
Total time slots & $20000$\\
Precoder & Zero-forcing\\
Channel model & i.i.d. Rayleigh fading model\\
\hline
\end{tabular}
\end{table}
\begin{figure}[!t]
\centering
\includegraphics[width=0.4\textwidth]{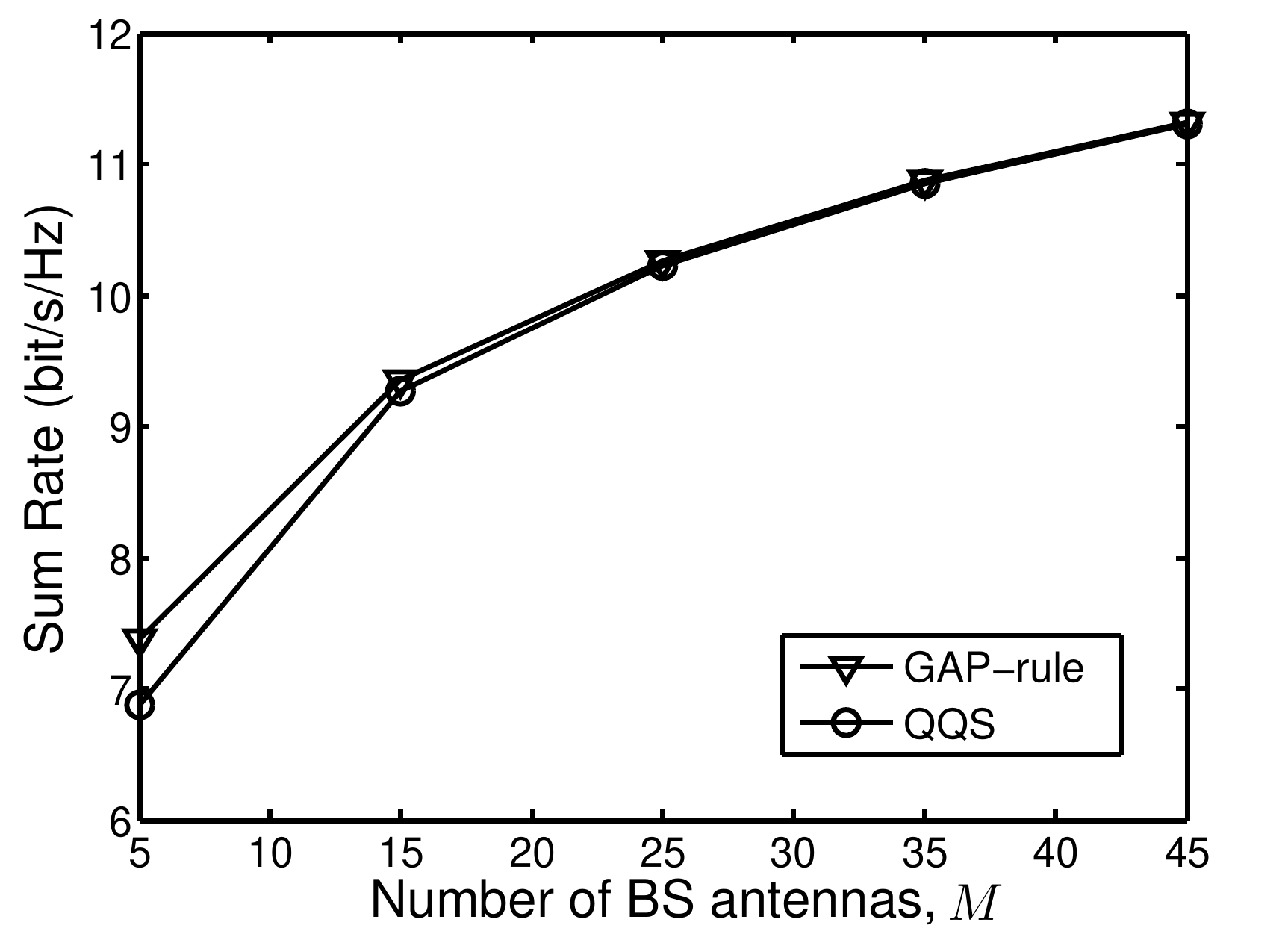}
\caption{Cell sum rate of GAP-rule and QQS with user channel coherence time given in Table \ref{UCTa}. The number of users $N=5$. The number of user-groups is $K=2$. }
\label{Fig_QQSQ}
\end{figure}
\begin{figure}[!t]
\centering
\includegraphics[width=0.4\textwidth]{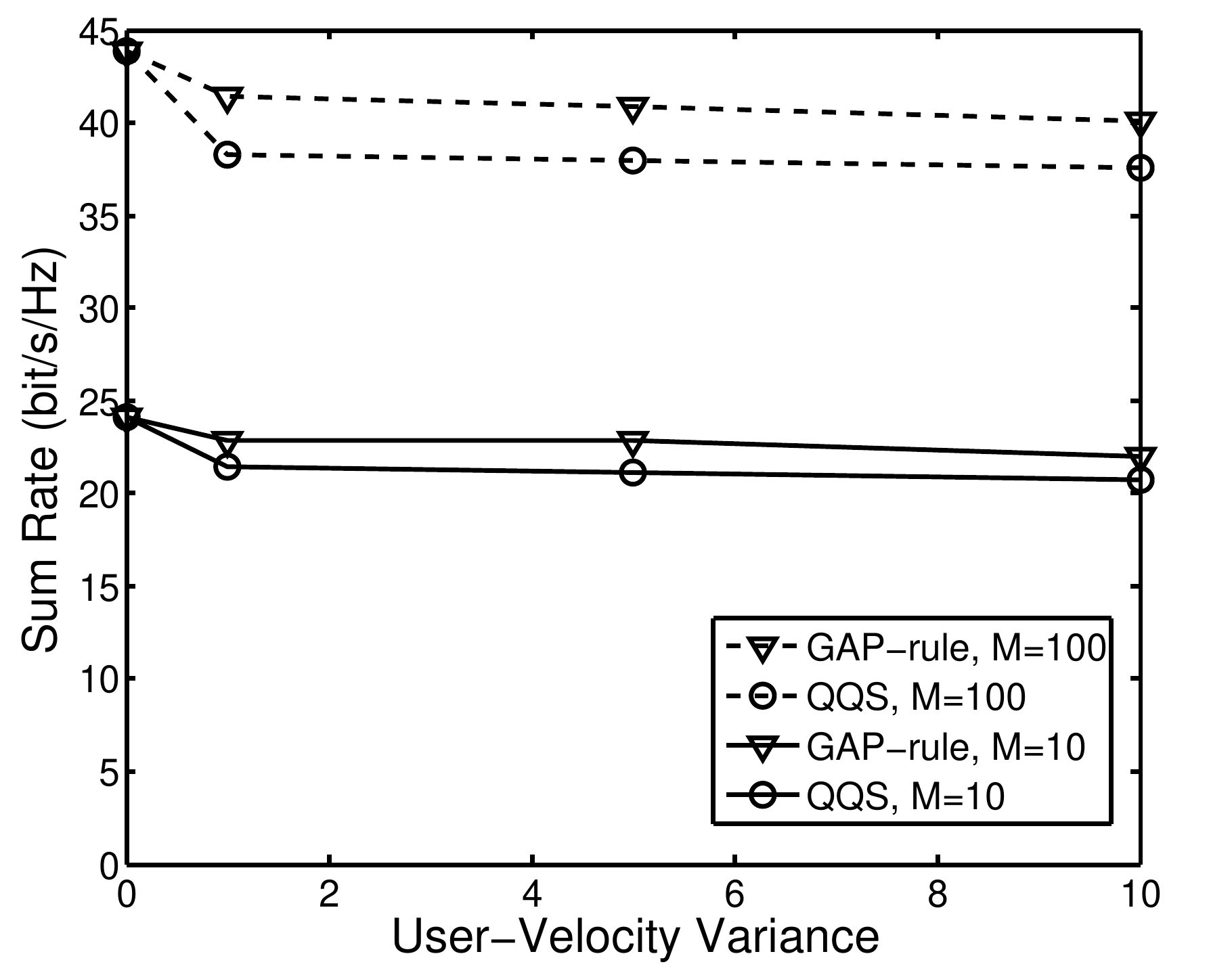}
\caption{Cell sum rate of GAP-rule and QQS with Gaussian random user-velocity. The number of users $N=10$. The number of user-groups, i.e., K, is optimized by exhaustive search.}
\label{Fig_QQSQCT}
\end{figure}
\begin{figure}[!t]
\centering
\includegraphics[width=0.4\textwidth]{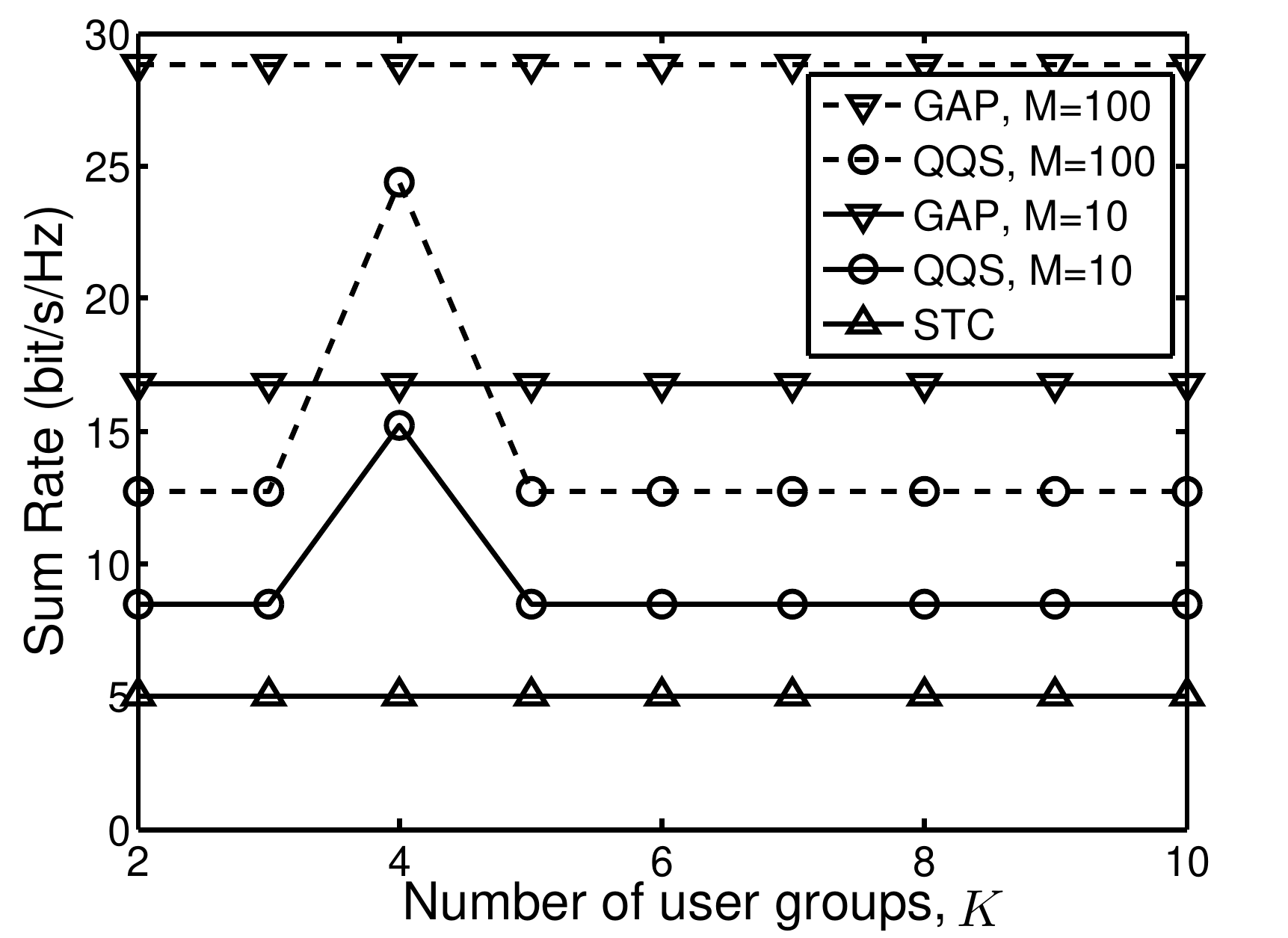}
\caption{Cell sum rate of GAP-rule and QQS with uniformly-distributed random user-velocity. The number of users $N=10$. }
\label{Fig_QQSK}
\end{figure}
\section{Numerical Results}
\label{NR}
In this section, we show the performance of our proposed schemes through computer simulations. The parameters used in the simulations are shown in Table \ref{tab_para}. First, to illustrate the performance of the QQS, we compare the QQS with the throughput-optimal GAP-rule. The sub-optimality of the QQS is shown, which stems from the fact that the QQS heuristically makes two simplifications of the GAP-rule, namely ignoring the channel fluctuations and grouping the users based on channel coherence times. In Fig. \ref{Fig_QQSQ}, we demonstrate the sub-optimality due to neglecting channel fluctuations, by letting the user channel coherence times be naturally grouped as shown in Table \ref{UCTa}, thus eliminating the sub-optimality due to user grouping. We consider a scenario where two types of users coexist: $2$ high-mobility ($60$ km/h) and $3$ low-mobility ($3$ km/h) users. The user channel coherence times\footnote{Note that we refer to the channel coherence time as the block length in the block fading model in this paper. Because users in one cell usually have identical channel coherent bandwidths, different block lengths of users are mainly due to different coherence times. Therefore we use channel coherence time instead of block length in the paper for better illustration.} in terms of the number of channel uses are shown in Table \ref{UCTa},
\begin{table}[!t]
\renewcommand{\arraystretch}{1.3}
\caption{User coherence times}
\centering
\begin{tabular}{|c |c |c |c |c|}
\hline
User $1$ & User $2$ & User $3$ & User $4$ & User $5$\\
\hline
100&100&100&5&5\\
\hline
\end{tabular}
\label{UCTa}
\end{table}
according to
\begin{equation}
T = B_\textrm{c} \times T_\textrm{c},
\end{equation}
where $B_\textrm{c}=\frac{c}{4\Delta d}$ and $T_\textrm{c}=\frac{1}{8f_c vc}$, $c$ denotes the light speed, $\Delta d$ is related to the cell radius and $v$ denotes the user velocity \cite{Tse05}. We run the simulation of the algorithms for $20000$ time slots, which is $1.3$ seconds under these parameters, and compute the sum rate by averaging the service rate based on Corollary \ref{Thm_TR}. It is observed that the QQS is asymptotically throughput-optimal in the large-system regime. Despite of the sub-optimality when the number of BS antennas is limited, the rate loss is marginal shown in the figure, e.g., when $M=5$, the sum rate loss is approximately $0.5$~bit/s/Hz. The marginal rate loss is because that it is well-known the user-rates with linear beamforming converge to the so-called deterministic equivalents quite fast, as the system dimension goes up \cite{Wagner12}. Therefore it is reasonable for the QQS to put aside the channel fluctuations and focus on the queue information. On the other hand, since the T-DCA and PL-DCA schemes are both throughput-optimal, the sum rate plots for them coincide with the GAP-rule.

Fig. \ref{Fig_QQSQCT} shows the QQS performance when the user coherence times are random, in order to investigate the sub-optimality due to user grouping. We let the user-velocities be truncated Gaussian distributed with means being $3$ and $60$ km/h\footnote{The negative velocities are eliminated and re-generated. }, and the variance $\sigma_\textrm{v}^2$ is given as the x-axis of the figure. It is observed that when $\sigma_\textrm{v}^2=0$, i.e., the channel coherence times are naturally grouped, the QQS performs as good as the GAP-rule, confirming the intuition of avoiding scheduling users with dramatically distinct channel coherence times. Furthermore, when the user-velocities vary, the rate loss of the QQS is fairly acceptable, given the fact that the QQS not only dramatically decreases the complexity, but also makes the algorithm practical comparing with the GAP-rule. It is also observed that the rate gap of $M=100$ is larger than that of $M=10$, whereas relatively, the rate gaps are similar given the relative difference. This implies that the analytic expression of the rate gap may involve a term that scales with $M$, possibly as $\log(M)$ since this is the form of the power gain.

\begin{figure}[!t]
\centering
\includegraphics[width=0.4\textwidth]{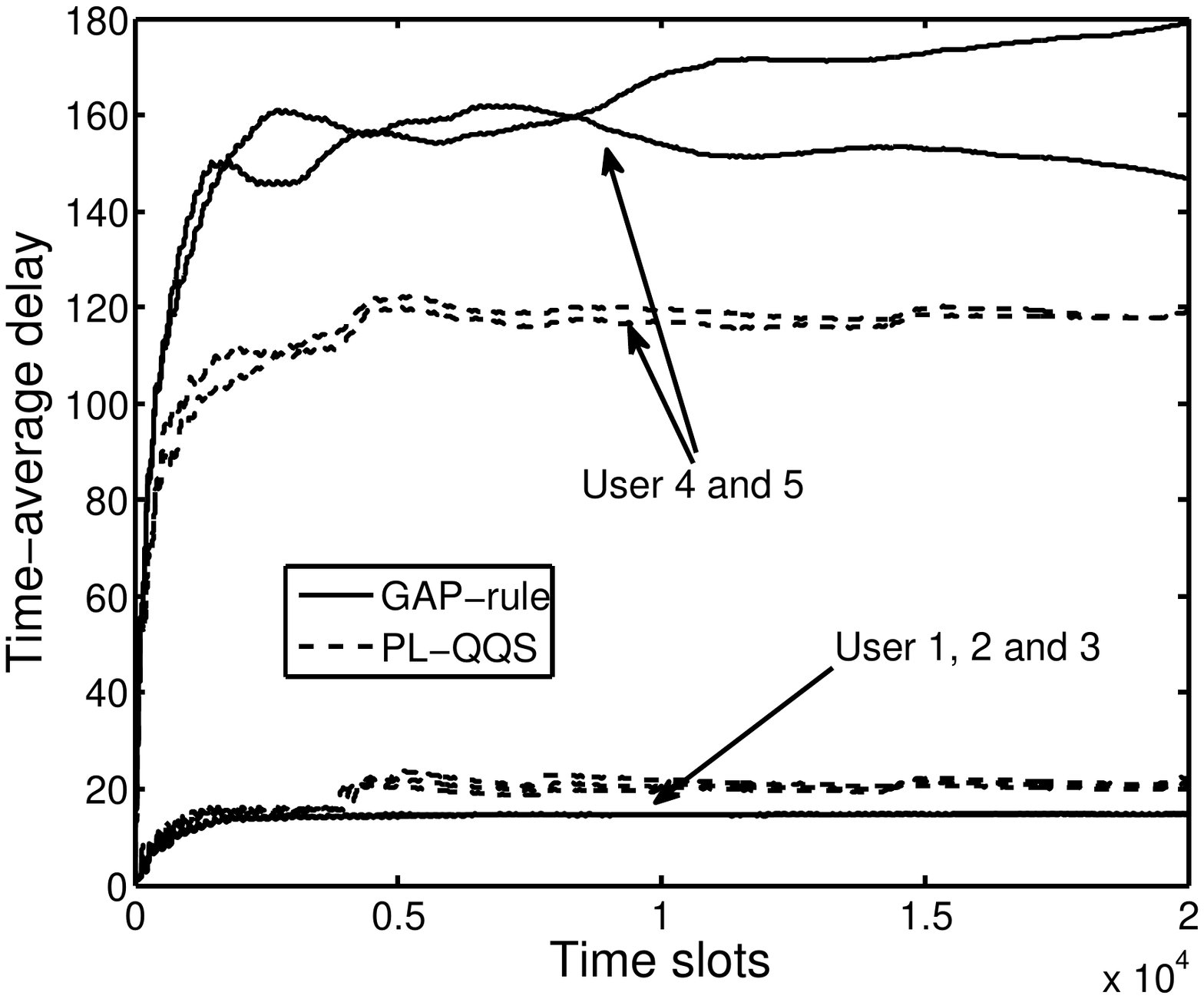}
\caption{Time-average delay performance of the GAP-rule.}
\label{Fig_MWT}
\end{figure}
\begin{figure}[!t]
\centering
\includegraphics[width=0.4\textwidth]{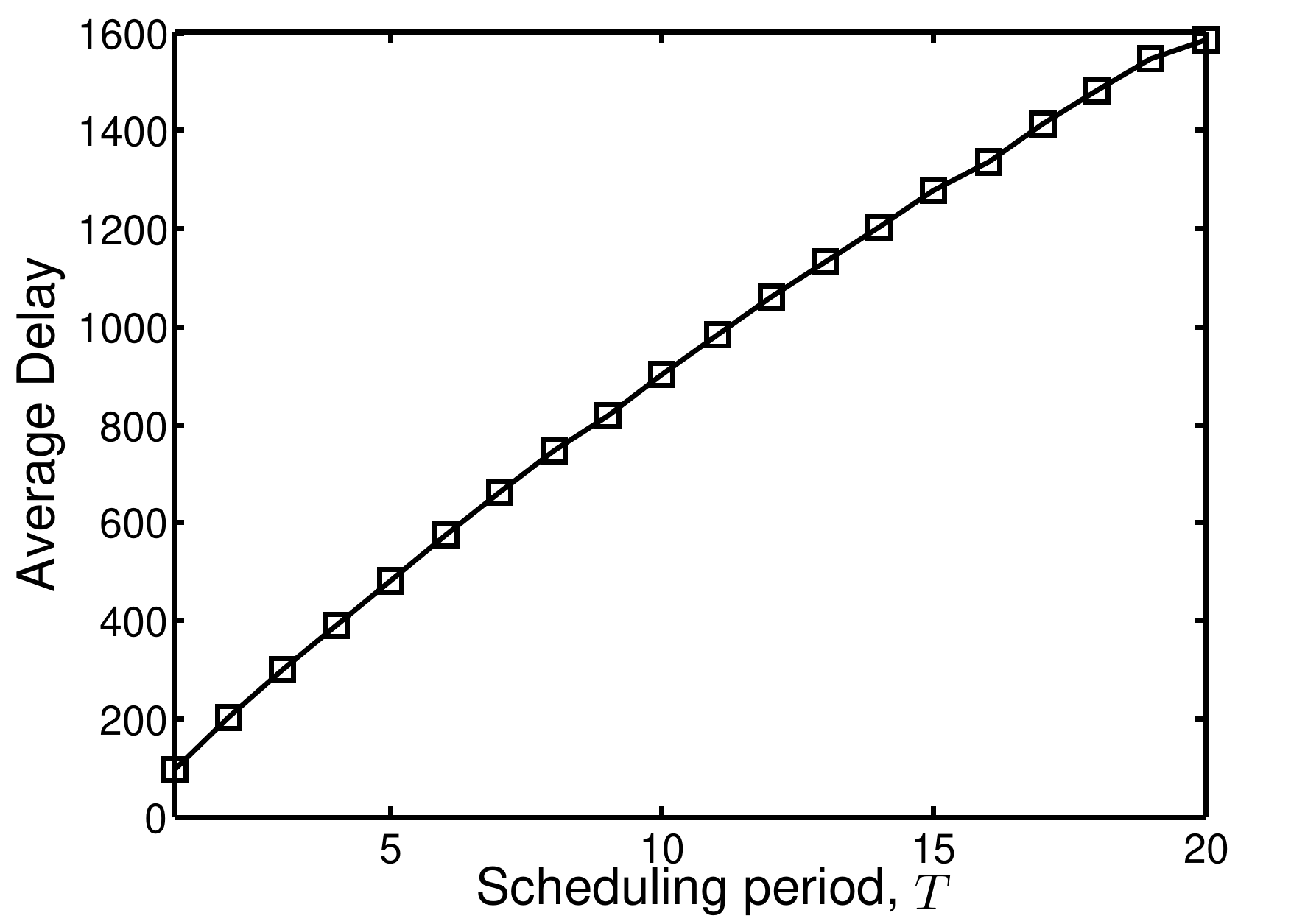}
\caption{Average delay performance of the T-DCA scheme.}
\label{Fig_T}
\end{figure}
\begin{figure}[!t]
\centering
\includegraphics[width=0.4\textwidth]{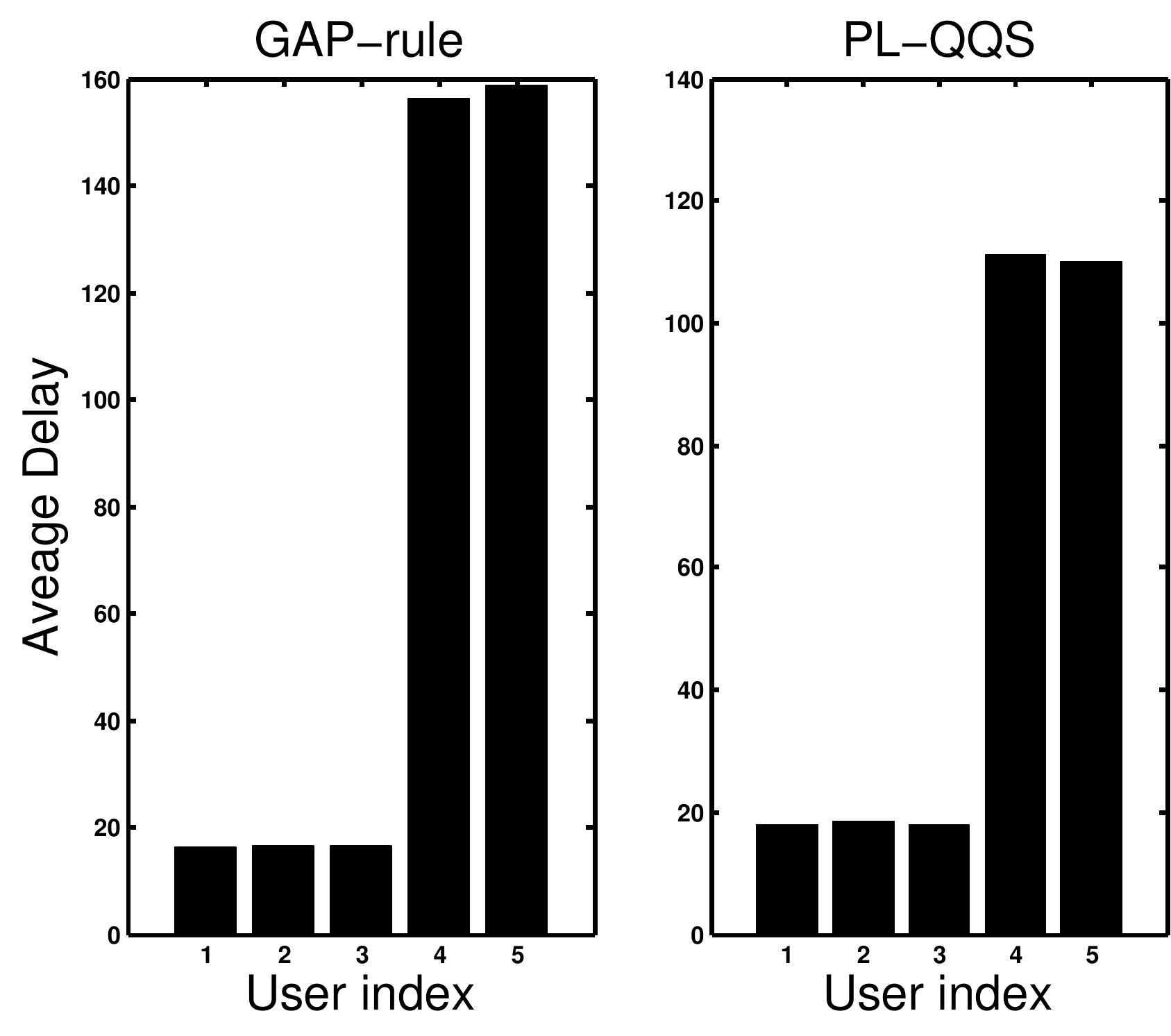}
\caption{Average delay comparison between GAP-rule, and PL-QQS schemes, where $\theta=3$.}
\label{Fig_DF}
\end{figure}
The impact of the number of user-groups in the QQS is shown in Fig. \ref{Fig_QQSK}. It is important to set the number of user groups $K$ in the QQS, since the QQS only allows transmission to users in the same group exclusively. Specifically, the channel coherence time approximation in each group will be inaccurate when $K$ is too small, i.e., the number of users in each group is too large. On the other hand, over-grouping the users, i.e., large $K$, simply leads to time-sharing among different users. It is observed that there exists an optimum $K$, with which the performance of the QQS is fairly close to the GAP-rule. The analytic analysis of the optimum $K$ is not given due to the heuristics of the QQS algorithm and the difficulty to analyze it. However, given the limited searching space of $K$, which at most scales linearly with the number of users, and the fact that the search is only required as often as the user channel coherence time changes, which is shown to be on the order of seconds to tens of seconds, an exhaustive search is acceptable. Nonetheless, the exact analysis is left to be an interesting future work. Note that the design of better user-grouping schemes, rather than uniform and fixed quantization of the channel coherence times as in \eqref{CHTQ}, is also worth studying in the future.

For comparison purposes, the sum rate of simply time-division-multiple-access (TDMA) among users is also plotted, which is evidently inferior compared to QQS or GAP-rule. Remark that multiplexing all users generates a \emph{zero} throughput, given the user coherence times in this simulation, due to the CSIT acquisition overhead occupies all the available time-frequency resources.

Fig. \ref{Fig_MWT} shows the time-average delay performance for the GAP-rule and the PL-QQS scheme. The parameters are identical with those in Fig. \ref{Fig_QQSQ} and $M=10$. The arrival processes are independent Binomial processes with arrival rates $\lambda_n = 1.5$.  The time-average delay at time $t$ of user $n$ is defined as
\begin{equation}
\label{TA-delay}
\bar D_n(t) = \frac{1}{t}\sum\limits_{\tau  = 1}^t { {{D_n}(\tau )} } ,
\end{equation}
where $D_n(\tau)$ denotes the head-of-line packet delay of Queue-$n$ \cite{Neely14}. From Fig. \ref{Fig_MWT}, we can observe that the system is stabilized as the queue size does not ``blow up''. It is shown that the delay-fairness is improved by the PL-DCA rule compared with the GAP-rule. The delay performance of the $T$-frame scheme is shown in Fig. \ref{Fig_T}, where we can observe the average delay is approximately \emph{linear} with the parameter $T$, which decides how often the user-selection is changed, while maintaining the stability. In practice, $T$ must be chosen such that the delay performance can satisfy the quality-of-service (QoS) requirement and also relieve the burden of rapidly modifying the frame structure. Note that the delay difference between T-DCA and the GAP-rule is due to the increased scheduling decision interval introduced by the T-DCA scheme, whereas the T-QQS scheme does not introduce more delay compared to the T-DCA, except the inherent performance degradation due to the heuristics of the QQS-based schemes, which is shown in Fig. \ref{Fig_QQSQ}-\ref{Fig_QQSK}.

Fig. \ref{Fig_DF} shows the user average delay performance of the proposed schemes. In the PL-QQS scheme, $\theta$ is set to be $3$. It turns out that the PL-QQS scheme achieves better delay-fairness than the GAP-rule in the sense of the average delay of the ST users are decreased substantially. In the mean time, the average delay of the LT users is not ``dragged up'' because we are actually utilizing the resources saved by avoiding ``over-service''. Note that while the delay-fairness of the users is improved, the throughput-optimality is still valid for the PL-DCA scheme.
\section{Conclusions}
\label{sec_c}
In this paper, we have investigated the user-scheduling problem in MU-MIMO downlinks considering CSIT acquisition overhead. It was shown that the performance of a system regardless of the CSIT acquisition overhead is very poor when the channel coherence time is comparable with the CSIT acquisition overhead. Therefore, a CSIT-overhead aware user scheduling scheme is in great need. We found that the Lyapunov-drift optimization can be leveraged to design a throughput-optimal user scheduling rule in MU-MIMO downlinks, namely the GAP-rule. Furthermore, by scheduling users at a slower rate based on the same GAP-rule, the T-DCA scheme is still throughput-optimal and we can strike a good balance between user delay performance and signaling overhead. In addition, considering the user delay fairness, it was found that a modified GAP-rule, i.e., the PL-DCA, based on which a larger power of the queue term in the GAP-rule is prescribed, achieves better user delay fairness while maintaining throughput-optimality.

Moreover, we designed the QQS-based schemes to realize the GAP-rule in practice and to significantly reduce the complexity. It was shown that the QQS performs fairly close to the GAP-rule, when the system dimension is large and the user coherence times are naturally grouped. The QQS suffers reasonable rate loss when either condition is not met exactly. Nevertheless, the performance improvement comparing with full spatial multiplexing or simple TDMA is evident.

To apply our proposed schemes, one should notice that they are designed for TDD MU-MIMO systems. Whereas in FDD systems, the channel estimations for uplink and downlink are decoupled, therefore, it is possible to schedule different set of users in the uplink and downlink. The user scheduling in this scenario requires special treatment. In addition, recent work on exploiting second-order channel statistics (SOCS), e.g., \cite{Adhikary13}\cite{Jiang14}, has shown that virtual sectorizations bring a new dimension to the user scheduling problem, in the sense that by leveraging the SOCS, the system can schedule users simultaneously with distinctly disjoint angular spreads without instantaneous CSIT. Both topics, namely the user scheduling for FDD systems and with virtual sectorizations, are very intriguing topics and worth research attention in the future.

\appendices
\section{Proof of Theorem \ref{Thm_TO}}
\label{proof_thm1}
We first briefly review the Lyapunov-drift approach, which is the workhorse in our proof. Define the Lyapunov function as
\begin{equation}
\label{Lfunction}
L(t_k) \triangleq \frac{1}{2}\sum\limits_n {Q_n^2({t_k})},
\end{equation}
and the Lyapunov drift as
\begin{equation}
\label{Ldrift}
\Delta (t_k)\triangleq \mathbb{E}[L(t_{k+1})-L(t_k)|\bm{Q}(t_k)].
\end{equation}
\newtheorem{lemma}{Lemma}
\begin{lemma}
\label{lemma1}
If there exist constants $B$ and $\epsilon$, which satisfy
\begin{equation}
\label{dstability}
\Delta (t_k) \le B -\epsilon \sum\limits_n {Q_n({t_k})} T_k,
\end{equation}
where $T_k = t_{k+1}-t_k$, then we have
\begin{equation}
\label{QueueStk}
\mathop {\lim \sup }\limits_{T \to \infty } \frac{{\sum\limits_{k = 1}^T  \mathbb{E}[\sum\limits_n {{Q_n}({t_k})} ]{T_k}}}{{\sum\limits_{k = 1}^T {{T_k}} }} < \infty ,
\end{equation}
and all queues are strongly stable.
\end{lemma}
\begin{IEEEproof}
Lemma \ref{lemma1} is slightly different from the strong stability defined in \cite{Neely10}, wherein the queue size is defined on the whole time domain. The key notion is that leveraging the boundedness of the arrival rates and service rates, the condition \eqref{QueueStk} can be written in the form of \eqref{QueueS}, with the difference bounded above by a constant.
\end{IEEEproof}
Given the queuing dynamics \eqref{QD}, we have
\begin{IEEEeqnarray}{lcl}
\label{thmp1}
\Delta ({t_k}) &\le& \mathbb{E}\left[\sum\limits_n \left. {\frac{{\beta _n^2({t_k}) + \alpha _n^2({t_k})}}{2}} \right|\bm{Q}({t_k})\right] \nonumber\\
&& - \sum\limits_n {{Q_n}({t_k})} \mathbb{E}\left[{\beta _n}({t_k}) - {\alpha _n}({t_k})|\bm{Q}({t_k})\right].
\end{IEEEeqnarray}
Observing that
\begin{IEEEeqnarray}{lcl}
\label{thmp2}
\mathbb{E}&&\left[\sum\limits_n \left.{\frac{{\beta _n^2({t_k}) + \alpha _n^2({t_k})}}{2}} \right|\bm{Q}({t_k})\right] \nonumber\\
&& \le \frac{1}{2}\left[r_{\textrm{max}}+ A_{\textrm{max} }^2 \right]{T_{\textrm{max}}^2} \triangleq B,
\end{IEEEeqnarray}
where $r_{\textrm{max}}$ and $T_{\textrm{max}}$ denote the maximum service rate and the maximum frame length, respectively. Consider the following optimization problem
\begin{IEEEeqnarray}{lcl}
\label{RP}
 \textrm{maximize:}&\quad&  \epsilon \nonumber \\
  \textrm{subject to:} &\quad& \bar \mu_n = \sum\limits_{s = 1}^S {{P_{{\pi _s}}}\frac{{{\beta _{\pi _s,n}}}}{{{T_{\pi_s}}}}}  \ge {\lambda _n} + \epsilon, \,\forall n,
\end{IEEEeqnarray}
where $\pi_s$ denotes any scheduling action which is feasible, $\beta _{\pi _s,n}$ and $T_{\pi_s}$ denote the allocated service bits during the frame and frame length under the control policy $\pi_s$, respectively, and $P_{{\pi _s}}$ denotes the probability of taking the action $\pi_s$. The solution of this problem will lead us to a \emph{randomized policy} $\omega^*$, according to which, we take the action $\pi_s$ with probability $P_{{\pi _s}}$ at the beginning of each frame. $\bar \mu_n$ is the time-average allocated service rate. Based on \cite{Neely10}, under some mild conditions (ergodicity for example), any achievable ergodic rate point $\bm{\Lambda}  = ({\lambda _1},{\lambda _2},...,{\lambda _n})$ can be achieved by $\omega^*$. Plugging this randomized policy $\omega^*$, which is independent with $\bm{Q}({t_k})$, into \eqref{thmp1}, we obtain
\begin{IEEEeqnarray}{lcl}
  \Delta ({t_k}) &\le& B - \sum\limits_n {{Q_n}({t_k})} \mathbb{E}\left[ {{\beta _n}({t_k}) - {\alpha _n}({t_k})|{\bm{Q}}({t_k})} \right]\nonumber \\
   &=& B - \nonumber\\
   \label{dr1}
   &&\sum\limits_n {{Q_n}({t_k}){T_k}} \mathbb{E}\left\{\left[ {\left. \frac{{{\beta _n}({t_k})}}{{{T_k}}}\right|{\bm{Q}}({t_k})} \right] - {\lambda _{\rm{n}}}\right\} \\
   \label{dr2}
   &\le& B - \sum\limits_n {{Q_n}({t_k}){T_k}} [{{\bar \mu }_{\rm{n}}} - {\lambda _{\rm{n}}}]  \\
   \label{dr3}
   &\le& B - \epsilon \sum\limits_n {{Q_n}({t_k}){T_k}},
\end{IEEEeqnarray}
where equality \eqref{dr1} bases on the fact that the arrival process is i.i.d. and independent of the queue size, inequality \eqref{dr2} comes from the definition of \eqref{MWTR}, and by the constraint in \eqref{RP}, we have inequality \eqref{dr3}. And based on Lemma \ref{lemma1}, we complete the proof.
\section{Proof of Corollary \ref{coro1}}
\label{proof_thm3}
The T-DCA falls into the category of imperfect scheduling defined in \cite{Neely10}, based on which, it is sufficient to prove the Lyapunov-drift under this scheme is within a constant to the optimum. Note that the optimal scheduling \eqref{MWTR} chooses the users every frame instead of every $T$ frames. Consider one drift of the T-DCA policy. Let $\{\delta_1,\delta_2,...,\delta_f\}$ be the optimal scheduling frame-length during this drift interval $T_f$, where
\begin{equation}
{T_f} \approx \sum\limits_i {{\delta _i}},
\end{equation}
neglecting the time that may exist when the drift interval does not contain these $f$ frames exactly, assuming $T_f$ is relatively large. Let $\beta_n^*(i)$ and $\beta_n(T)$ be the optimal allocated service bits during time $\delta_i$ and the T-DCA service bits during the drift interval $T_f$, respectively, where $t_i$ is the time at the beginning of time interval $\delta_i$. We have
\begin{IEEEeqnarray}{lcl}
&& \sum\limits_n {{Q_n}({t_i})} \frac{{\beta _n^*(i)}}{{{\delta _i}}} \nonumber\\
&\le& \sum\limits_n {({Q_n}({t_1}) + {A_{\max }}{(t_i-t_1)})} \frac{{\beta _n^*(i)}}{{{\delta _i}}} \nonumber \\
\label{thm3_1}
&\le& \sum\limits_n {{Q_n}({t_1})} \frac{{{\beta _n}(T)}}{{{T_f}}} + \sum\limits_n {{A_{\max }}{(t_i-t_1)}} \frac{{\beta _n^*(i)}}{{{\delta _i}}} \\
\label{thm3_2}
&\le& \sum\limits_n {{Q_n}({t_1})} \frac{{{\beta _n}(T)}}{{{T_f}}} + N T{A_{\max }} \max{[T_n]}{R_{\max }},
\end{IEEEeqnarray}
where (\ref{thm3_1}) follows from the optimality of the T-DCA given the queue information at the time $t_1$, and \eqref{thm3_2} follows because the service rate and the time interval can both be upper-bounded by constants. Therefore the drift of the T-DCA can be bounded below by the optimum subtracts a constant, which concludes the proof.
\section{Proof of Theorem \ref{Thm_PLMW}}
\label{proof_thm2}
For an odd integer $\theta$, based on the fact that for any $a \in \mathbb{R}, (\max[a,0])^{\theta+1} \le a^{\theta+1}$, we have
\begin{IEEEeqnarray}{rcl}
\label{thm2_1}
\frac{1}{\theta }\sum\limits_n {Q_n^{\theta  + 1}({t_{k + 1}})}  && \le \frac{1}{\theta }\sum\limits_n {Q_n^{\theta  + 1}({t_k})} - \sum\limits_n {Q_n^\theta ({t_k})} (\beta_n ({t_k}) \nonumber\\
&& - \alpha_n ({t_k})) + \sum\limits_n {o(Q_n^\theta ({t_k}))},
\end{IEEEeqnarray}
where $\smallO{f^{\theta}(x)}$ denotes a polynomial of $f(x)$ which has a lower order than $\theta$. Define the Lyapunov function as
\begin{equation}
\label{Lt}
L(t_k) \triangleq \frac{1}{\theta+1}\sum\limits_n {Q_n^{\theta+1}({t_k})},
\end{equation}
combining with \eqref{thm2_1}, the Lyapunov-drift is
\begin{IEEEeqnarray}{rcl}
\label{thm2_2}
{\Delta ({t_k}}) && \le  - \sum\limits_n {Q_n^\theta ({t_k})}T_k \mathbb{E}\left[\frac{\beta ({t_k})}{T_k} - \lambda_n|\bm{Q}(t_k)\right] \nonumber\\
&& + \sum\limits_n {\smallO{Q_n^\theta ({t_k})}}.
\end{IEEEeqnarray}
\begin{lemma}
\label{lemma2}
There exists an $M<\infty$, such that for any $\bm{Q}(t_k)$ satisfying $L(t_k)>M$, we have
\begin{equation}
\Delta_P(t_k)<0,
\end{equation}
where $\Delta_P(t_k)$ denotes the Lyapunov-drift under the PL-DCA.
\end{lemma}
\begin{IEEEproof}
Plug in the randomized policy of \eqref{RP}, we have
\begin{equation}
\label{delta}
\Delta_P(t_k)\le - \sum\limits_n {Q_n^\theta ({t_k})} {T_k}\epsilon  + \sum\limits_n {\smallO{Q_n^\theta ({t_k})}}.
\end{equation}
Observing that
\begin{equation}
\label{order_2}
\sum\limits_n {Q_n^\theta ({t_k})} \ge Q_{n^*}^\theta ({t_k}),
\end{equation}
where
\begin{equation}
\label{order_1}
n^*=\textrm{argmax}_n  (Q_n(t_k)),
\end{equation}
and that an expression which is of the order ${\smallO{Q_n^\theta ({t_k})}}$ can be bounded above by ${\smallO{Q_{n^*}^\theta ({t_k})}}$ according to \eqref{order_1}, we have
\begin{equation}
\Delta_P(t_k)\le - Q_{n^*}^\theta ({t_k}) {T_k}\epsilon  + \sum\limits_n {\smallO {Q_{n^*}^\theta ({t_k})}}.
\end{equation}
Since
\begin{equation}
\label{LL}
L({t_k}) \le \frac{1}{{\theta  + 1}}NQ_{{n^*}}^{\theta  + 1}({t_k}),
\end{equation}
then
\begin{equation}
Q_{{n^*}}^{\theta  + 1}({t_k}) \ge \sqrt[{\frac{1}{{\theta  + 1}}}]{{\frac{{(\theta  + 1)L({t_k})}}{N}}}
\end{equation}
follows. Observing the order of $Q_n ({t_k})$ on the right-hand-side of equation \eqref{delta}, there exists $0<\omega<\infty$, such that for any
$Q_{n^*}^{\theta+1} ({t_k})>\omega$, we have $\Delta_P(t_k)<0$. To ensure this, let
\begin{equation}
M>\frac{1}{{\theta  + 1}}N\omega^{\theta  + 1},
\end{equation}
combining with \eqref{LL}, we conclude the proof of the Lemma.
\end{IEEEproof}

Now we apply Lemma \ref{lemma2} to prove an upper bound for $L(t_k)$. To this end, suppose $L(t_k)\le M$, then
\begin{IEEEeqnarray}{lcl}
L(t_{k+1})&\le&\frac{1}{{\theta  + 1}}\sum\limits_n {{{({Q_n}({t_k}) + {A_{\max }})}^{\theta  + 1}}} \nonumber\\
&=& \frac{1}{{\theta  + 1}}\sum\limits_n {\sum\limits_i {\binom{\theta+1 }{i}A_{\max }^iQ_n^{\theta  + 1 - i}({t_k})} } \nonumber \\
&\le& \frac{1}{{\theta  + 1}}\sum\limits_n {\sum\limits_i {A_{\max }^iQ_n^{\theta  + 1}({t_k})} } \nonumber \\
&\le& \sum\limits_i^{\theta+1} {A_{\max }^i} M.
\end{IEEEeqnarray}
On the other hand, suppose $L(t_k)>M$, we have $\mathbb{E}\left[L(t_{k+1})\right]<\mathbb{E}\left[L(t_{k})\right]$ based on Lemma \ref{lemma2} together with taking the iterative expectation. Therefore $\mathbb{E}\left[L(t_k)\right]\le\sum\limits_i^{\theta+1} {A_{\max }^i} M$. Combining with the \emph{Jensen's} inequality, we have
\begin{IEEEeqnarray}{lcl}
\sum\limits_n {\mathbb{E}[{Q_n}({t_k})}]  &\le& N\left[{\frac{{\theta  + 1}}{N}\mathbb{E}[L({t_k})]}\right]^{\frac{1}{1+\theta}} \nonumber \\
&\le& N\left[{\frac{{\theta  + 1}}{N}{\sum\limits_i {A_{\max }^i} M}}\right]^{\frac{1}{1+\theta}}.
\end{IEEEeqnarray}
Dividing both sides by $t_k$ and let $t_k$ go to infinity, we can prove the system is mean-rate stable, which is a weaker, but sufficient stable condition for the system.
\bibliographystyle{ieeetr}
\bibliography{4}
\end{document}